\PassOptionsToPackage{usenames, table, dvipsnames}{xcolor} 

\documentclass[lettersize,journal]{IEEEtran}

\def\BibTeX{{\rm B\kern-.05em{\sc i\kern-.025em b}\kern-.08em
    T\kern-.1667em\lower.7ex\hbox{E}\kern-.125emX}}
\usepackage{balance}

\def\plotFlag{true} 

\usepackage[footnotesize,labelfont=bf]{caption}

\usepackage{orcidlink}

\usepackage{wrapfig}

\usepackage{cite}
\usepackage{amsmath,amssymb,amsfonts}
\usepackage{nicefrac, xfrac}
\usepackage{graphicx}
\usepackage{textcomp}
\usepackage{xstring} 

\usepackage{enumerate}

\usepackage{xcolor}

\usepackage{amsmath, bm}
\usepackage{mathtools} 
\DeclarePairedDelimiter{\ceil}{\lceil}{\rceil}
\usepackage{bigstrut}

\usepackage{subcaption}

\usepackage{cleveref}
\Crefname{figure}{Fig.}{Figs.}

\usepackage{siunitx}
\DeclareSIUnit\bit{b}
\DeclareSIUnit{\bps}{bps}

\usepackage[export]{adjustbox} 

\usepackage{pgf}
\usepackage{tikz}
\usetikzlibrary{shapes.geometric}
\usetikzlibrary{shapes.arrows}
\usetikzlibrary{arrows}
\usetikzlibrary{arrows.meta}
\usetikzlibrary{decorations.pathreplacing}
\usetikzlibrary{calc}
\def\centerarc[#1](#2)(#3:#4:#5)
    { \draw[#1] ($(#2)+({#5*cos(#3)},{#5*sin(#3)})$) arc (#3:#4:#5); }
\definecolor{apBlue}{RGB}{0,64,122}
\definecolor{ueRed}{RGB}{255, 0, 0}

\definecolor{hl_gray}{RGB}{160,160,160}
\definecolor{wr_gray}{RGB}{200,200,200}
\definecolor{pw_gray}{RGB}{240,240,240}

\usepackage{booktabs}
\usepackage{tabularx}
\newcolumntype{Y}{>{\centering\arraybackslash}X}
\usepackage{multirow}

\newcommand\clearrow{\global\let\rowmac\relax}
\clearrow
\usepackage{diagbox} 

\usepackage[nogroupskip, nopostdot]{glossaries} 
\glsdisablehyper 
\newacronym{ack}{ACK}{acknowledgement}
\newacronym{ampdu}{A-MPDU}{aggregated MAC protocol data unit}
\newacronym{amsdu}{A-MSDU}{aggregated \gls{msdu}}
\newacronym{aod}{AoD}{angle of departure}
\newacronym{aoi}{AoI}{age of information}
\newacronym{ap}{AP}{access point}
\newacronym{ar}{AR}{augmented reality}
\newacronym{a-bft}{A-BFT}{association beam forming training}
\newacronym{ber}{BER}{bit error ratio}
\newacronym{bft}{BFT}{beamforming training}
\newacronym{bf}{BF}{beamforming}
\newacronym{bi}{BI}{beacon interval}
\newacronym{bpc}{BPC}{bits per color}
\newacronym{brp}{BRP}{beam refinement period}
\newacronym{br}{BR}{beam refinement}
\newacronym{bss}{BSS}{basic service set}
\newacronym{bti}{BTI}{beacon transmission interval}
\newacronym{bt}{BT}{beam tracking}
\newacronym{cbap}{CBAP}{contention-based access period}
\newacronym{cots}{COTS}{consumer off-the-shelf}
\newacronym{cts}{CTS}{clear to send}
\newacronym{dmg}{DMG}{directional multi-gigabit}
\newacronym{dti}{DTI}{data transmission interval}
\newacronym{e2e}{E2E}{end-to-end}
\newacronym{evm}{EVM}{error vector magnitude}
\newacronym{fov}{FoV}{field of view}
\newacronym{fps}{FPS}{frames per second}
\newacronym{gi}{GI}{guard interval}
\newacronym{hmd}{HMD}{head-mounted display}
\newacronym{hpbw}{HPBW}{half power beamwidth}
\newacronym{ifs}{IFS}{inter-frame spacing}
\newacronym{imu}{IMU}{inertial measurement unit}
\newacronym{ip}{IP}{internet protocol}
\newacronym{kpi}{KPI}{key performance indicator}
\newacronym{lan}{LAN}{local area network}
\newacronym{ldpc}{LDPC}{low-density parity-check}
\newacronym{los}{LoS}{line of sight}
\newacronym{nlos}{NLoS}{non-line-of-sight}
\newacronym{olos}{OLoS}{obstructed \gls{los}}
\newacronym{m2p}{M2P}{motion-to-photon}
\newacronym{mcs}{MCS}{modulation and coding scheme}
\newacronym{mmw}{mmWave}{millimeter-wave}
\newacronym{mpc}{MPC}{multipath component}
\newacronym{mpdu}{MPDU}{MAC protocol data unit}
\newacronym{msdu}{MSDU}{MAC service data unit}
\newacronym{phy}{PHY}{physical layer}
\newacronym{ppdu}{PPDU}{PHY protocol data unit}
\newacronym{psdu}{PSDU}{PHY service data unit}
\newacronym{qoe}{QoE}{quality of experience}
\newacronym{aqoe}{AppQoE}{application quality of experience}
\newacronym{qos}{QoS}{quality of service}
\newacronym{rat}{RAT}{radio access technology}
\newacronym{rss}{RSS}{received singla strength}
\newacronym{rtp}{RTP}{real-time transport protocol}
\newacronym{rtxss}{\mbox{R-TXSS}}{responder transmit sector sweep}
\newacronym{rx}{RX}{receiver}
\newacronym{sls}{SLS}{sector-level sweep}
\newacronym{snr}{SNR}{signal-to-noise ratio}
\newacronym{sp}{SP}{service period}
\newacronym{ssw}{SSW}{sector sweep}
\newacronym{sta}{STA}{station}
\newacronym{stp}{STP}{setup message}
\newacronym{tof}{ToF}{time of flight}
\newacronym{tu}{TU}{time unit}
\newacronym{txss}{TXSS}{transmit sector sweep}
\newacronym{trn-t}{TRN-T}{training sequence}
\newacronym{trn-r}{TRN-R}{training sequence}
\newacronym{tru}{TRU}{training unit}
\newacronym{tx}{TX}{transmitter}
\newacronym{udp}{UDP}{user datagram protocol}
\newacronym{ue}{UE}{user equipment}
\newacronym{upa}{UPA}{uniform planar array}
\newacronym{uplink}{UL}{uplink}
\newacronym{ura}{URA}{uniform radial array}
\newacronym{vr}{VR}{virtual reality}
\newacronym{xr}{XR}{extended reality}
\newacronym{vr/ar}{VR/AR}{virtual/augmented reality}
\newacronym{thr}{THR}{throughput}
\newacronym{fbk}{FBK}{feedback}
\newacronym{me}{ME}{monitoring entity}
\newacronym{db}{DB}{database}
\newacronym{rl}{RL}{reinforcement learning}
\newacronym{soa}{SoA}{state-of-the-art}
\newacronym{mab}{MAB}{multi-armed bandit}
\newacronym{eirp}{EIRP}{equivalent isotropically radiated power}
\newacronym{cdf}{CDF}{cumulative distribution function}
\newacronym{ccdf}{CCDF}{complementary cumulative distribution function}

\newacronym{cir}{CIR}{channel impulse response}
\newacronym{ctf}{CTF}{channel transfer function}
\newacronym{mrc}{MRC}{maximum ratio combining}
\newacronym{hh}{HH}{horizontal-to-horizontal}
\newacronym{vv}{VV}{vertical-to-vertical}
\newacronym{agc}{AGC}{automatic gain control}
\newacronym{fspl}{FSPL}{free-space path loss}
\newacronym{ofdm}{OFDM}{orthogonal frequency-division multiplexing}
\newacronym{papr}{PAPR}{peak-to-average power ratio}
\newacronym{mimo}{MIMO}{multiple-input multiple-output}
\newacronym{fft}{FFT}{fast Fourier transform}
\newacronym{gpu}{GPU}{graphics processing unit}
\newacronym{de}{DE}{dominant eigenmode}

\newacronym{pdf}{PDF}{probability density function}

\newacronym{fdm}{FDM}{fully-digital \gls{mimo}}
\newacronym{det_full}{DET}{dominant eigenmode transmission}
\newacronym{det}{DET}{dominant eigenmode transmission}
\newacronym{sst}{SST}{spatial streaming transmission}
\newacronym{sm}{SM}{spatial multiplexing}
\newacronym{sage}{SAGE}{space alternating generalized expectation-maximization}
\newacronym{rf}{RF}{radio frequency}
\newacronym{eadf}{EADF}{effective aperture distribution function}
\newacronym{pmf}{PMF}{probability mass function}

\newacronym{isi}{ISI}{inter-symbol interference}
\newacronym{sc}{SC}{single carrier}
\newacronym{aoa}{AoA}{azimuth of arrival}
\newacronym{eoa}{EoA}{elevation of arrival}
\newacronym{gtd}{GTD}{geometrical theory of diffraction}
\newacronym{rms}{RMS}{root mean square}
\newacronym{pdp}{PDP}{power delay profile}
\newacronym{ple}{PLE}{path loss exponent}

\newcommand{\glsposs}[1]{%
  \ifglsused{#1}{%
    \glsdisp{#1}{\glsentryshort{#1}'s}%
  }{%
    \glsdisp{#1}{\glsentrylong{#1}'s (\glsentryshort{#1}'s)}%
  }%
}

\newcommand{\plotFig}[2]{
\IfEqCase{#1}{%
    {true}{#2}%
    {false}{\includegraphics[width=\linewidth]{alt_fig.png}}%
}[\PackageError{tree}{Undefined option to tree: #1}{}]%
}

\usepackage{algorithm}
\usepackage{algpseudocode} 

\newcommand{\vspacebelowfig}{\vspace{-0.5 cm}}

\usepackage{layouts} 

\usepackage[firstpage=true]{background}
\usepackage{eso-pic}

\backgroundsetup{
  position=current page.south,
  angle=0,
  nodeanchor=south,
  vshift=1mm,
  hshift=-2.6mm,
  opacity=1,
  scale=3,
  color=blue!80!white,
  contents={\begin{varwidth}{\textwidth}\fontsize{3}{6}\selectfont 
  © 2024 IEEE. Personal use of this material is permitted. Permission from IEEE must be obtained for all other uses, in any current \\
  or future media, including reprinting/republishing this material for advertising or promotional purposes, creating new collective \\
  works, for resale or redistribution to servers or lists, or reuse of any copyrighted component of this work in other works. \\
  This is the accepted version of the published paper with DOI: {\fontsize{4}{14}\selectfont 10.1109/TCOMM.2024.3392805}\\
  {\color{black}The measurement data is accessible at https://zenodo.org/records/10822725}
  \end{varwidth}
  }
}

\begin{document}

\title{Channel Performance Metrics and Evaluation for XR Head-Mounted Displays with mmWave Arrays}

\author{
Alexander~Marinšek~\orcidlink{0000-0001-9696-5365},~\IEEEmembership{Student Member,~IEEE,}
Xuesong~Cai~\orcidlink{0000-0001-7759-7448},~\IEEEmembership{Senior Member,~IEEE,}
Lieven~De~Strycker~\orcidlink{0000-0001-8172-9650},~\IEEEmembership{Member,~IEEE,}
Fredrik~Tufvesson~\orcidlink{0000-0003-1072-0784},~\IEEEmembership{Fellow,~IEEE,}
Liesbet~Van~der~Perre~\orcidlink{0000-0002-9158-9628},~\IEEEmembership{Senior Member,~IEEE}
\thanks{A. Marinšek, L. De Strycker, and L. Van der Perre are with \mbox{ESAT-WaveCore}, KU Leuven, Ghent, Belgium.
X. Cai and F. Tufvesson are with \mbox{LTH-EIT}, Lund University, Lund, Sweden.
Email of the corresponding authors: \{alexander.marinsek, liesbet.vanderperre\}@kuleuven.be}
}

\markboth{IEEE TRANSACTIONS ON COMMUNICATIONS,~Vol.~V, No.~N, MM~YYYY}%
{}

\maketitle

\begin{abstract}

\Gls{mmw} technology holds the potential to revolutionize \glspl{hmd} by enabling high-speed wireless communication with nearby processing nodes, where complex video rendering can take place. However, the sparse angular profile of \gls{mmw} channels, coupled with the narrow \gls{fov} of patch\nobreakdash-antenna arrays and frequent \gls{hmd} rotation, can lead to poor performance. 

We introduce six channel performance metrics to evaluate the performance of an \gls{hmd} equipped with \gls{mmw} arrays. We analyze the metrics using analytical models, discuss their impact for the application, and apply them to 28$\,$GHz channel sounding data, collected in a conference room using eight \gls{hmd} patch\nobreakdash-antenna arrays, offset by 45$^\circ$ from each other in azimuth. 

\indent Our findings confirm that a single array performs poorly due to the narrow \gls{fov}, and featuring multiple arrays along the \glsposs{hmd} azimuth is required. Namely, the broader \gls{fov} stabilizes channel gain during \gls{hmd} rotation, lessens the attenuation caused by \gls{los} obstruction, and increases the channel's spatial multiplexing capability. In light of our findings, we conclude that it is imperative to either equip the \gls{hmd} with multiple arrays or, as an alternative approach, incorporate macroscopic diversity by leveraging distributed \gls{ap} infrastructure.

\end{abstract}


\glsresetall

\section{Introduction}
A \gls{hmd} is a wearable piece of equipment that projects high\nobreakdash-fidelity visuals in front of the user's eyes. Rendering video on a nearby computer or edge processing node can reduce \gls{hmd} hardware complexity and aid ergonomics~\cite{FIROUZI2022101840}. \Gls{mmw} technology can provide the multi\nobreakdash-Gbps data rates required for \gls{xr} content streaming, owing to the ample bandwidth availability in the \gls{mmw} spectrum (\SIrange{30}{300}{\giga\hertz}) \cite{shafi_5g_2017, zhou_ieee_2018, THzMagazine}. To overcome large\nobreakdash-scale fading, \gls{mmw} transceivers feature directive antenna arrays. These augment the transceiver's gain to combat path loss and shadowing, while simultaneously mitigating the effects of small\nobreakdash-scale and frequency\nobreakdash-selective fading \cite{flordelis_spatial_2018}. However, \gls{mmw} channels are known for their sparsity. That is, they feature few powerful signal components. This begs the question: \emph{how many antenna arrays} should an \gls{hmd} have, and what are the \emph{relevant channel performance metrics} for \gls{xr} applications?

\Gls{mmw} channels are characterized by sparse multipath profiles. Analyses of channel measurement data, collected in office and conference room environments, show that \gls{mmw} channels seldom feature more than five \gls{mpc} clusters, among which only a few have sufficient gain and capturing the \gls{los} cluster is often essential for sustaining high\nobreakdash-throughput applications \cite{cai_dynamic_2020, gustafson_mm-wave_2014, wu_60-ghz_2017, huang_multi-frequency_2017, ju_millimeter_2021}. Hence, placing a patch\nobreakdash-antenna array with a \SI{-3}{\decibel} \gls{fov} of \SI{90}{\degree} in such a channel and having it continuously rotate on the head of an \gls{xr} user is akin to poor signal reception. 

In recent years, researchers have been investigating signal reception quality in the context of \gls{mmw} 5G NR mobile networks, and they have identified several adverse circumstances associated with handheld devices. These include movement, orientation changes, obstruction by objects, and physical contact between the device's antennas and the user. For instance, simply walking \SI{15}{\meter} along a corridor can lead to a degradation of signal strength by approximately \SIrange{15}{20}{\decibel}, with additional fluctuations of around $\pm$\SI{15}{\decibel} depending on the specific user \cite{hejselbak_measured_2017}. Changing the smartphone's orientation from vertical to horizontal can result in a halving of performance, as noted in \cite{aggarwal_first_2019}, which observed a \SI{50}{\percent} decrease in data rate in an IEEE~802.11ad network. Moreover, shadowing from the user's body can attenuate the signal by \SIrange{20}{25}{\decibel} in various handheld use cases \cite{zhao_user_2017, vaha-savo_empirical_2020}. Even holding the smartphone in one's palm can absorb up to \SI{15}{\decibel} of signal strength, while a mere finger interacting with the device can cause a \SI{3}{\decibel} disruption to the \gls{mmw} link \cite{xu_radiation_2018, xue_impacts_2023}. All these factors pose possible challenges to \gls{mmw} communications systems in \gls{xr} \glspl{hmd}. 

Previous studies on \gls{mmw} channel measurements and channel modeling for \glspl{hmd} have shown that high data rates can be achieved in cluttered indoor environments even with limited signal bandwidth \cite{venugopal_location_2016, gomes_mm_2018}, and that a single \gls{mmw} antenna array is likely insufficient \cite{struye_opportunities_2022}. Moreover, recent research advocates \gls{mmw}\nobreakdash-based sensing for complementing and even substituting the \glsposs{hmd} visual tracking systems \cite{taha_millimeter_2021, bedin_dopamine_2022}. Although prior art clearly outlines the potential and challenges associated with \gls{mmw} technology for \glspl{hmd}, there is a lack of insight into how broadening the \glsposs{hmd} \gls{fov} by employing multiple antenna arrays can alleviate the channel adversities that occur during \gls{hmd} rotation.

The work at hand conducts statistical and temporal characterization of \gls{mmw} channel gain and capacity for multi\nobreakdash-array \glspl{hmd}. Channel sounding data from a tailor\nobreakdash-made \SI{28}{\giga\hertz} measurement campaign is analyzed in order to assess the performance of different \gls{mmw} array configurations on the \gls{hmd}. Furthermore, we compare the results against analytical models and discuss their application\nobreakdash-level significance. Our contributions are summarized as follows:
\begin{itemize}
\vspace{0.5em}
\setlength\itemsep{0.5em}
    \item Derivation of six performance metrics, used for assessing the channel of a multi\nobreakdash-array \gls{hmd}, including their analysis using analytical models and a discussion of their application\nobreakdash-level significance,
    \item Analysis of the channel's performance in terms of the derived performance metrics for an \gls{hmd} employing \SIrange{1}{4}{}~\gls{mmw}~patch\nobreakdash-antenna arrays, and
    \item Evaluation of additional degrees of freedom, namely, the~usage of a rear\nobreakdash-\gls{hmd} headband for array placement, the deactivation of parts of the antenna arrays, and alternative \gls{ap} placement possibilities.
\vspace{0.5em}
\end{itemize}

We first describe the measurement campaign and post\nobreakdash-processing procedures in \Cref{sec:meas_setup_and_post_proc}. Then, we introduce the six performance metrics in \Cref{sec:performacne_metrics} and describe the analysis steps in \Cref{sec:analysis_procedures}. In \Cref{sec:res_antenna_array_configuration}, we report on the performance of configurations featuring \SIrange{1}{4}{}~arrays. In \Cref{sec:res_additional_considerations} we evaluate the effects of using a rear headband, altering the number of active \gls{hmd} antennas, and changing the \glsposs{ap} position. We summarize the main findings while providing an outlook for future research in \Cref{sec:conclusion}. Note that the paper at hand is an extension of our prior work, where we evaluated three of the six performance metrics \cite{marinsek_impact_2023}.

\emph{Notation and terminology:} $||\cdot||_\mathrm{F}$ is the Frobenius norm, $\mathbb{E}[\,\cdot\,]$~is the expectation operator, $e$ is the Euler number, and~$(\,\cdot\,)^{\text{T}}$ is the matrix transpose. The term \emph{gain} refers to the \emph{power gain} of the wireless channel, consisting of the combined channel and antenna response.

\section{Measurement setup and post-processing}
\label{sec:meas_setup_and_post_proc}

This section outlines the utilized channel sounder, the measurement environment and procedures, as well as the employed channel reconstruction and normalization techniques.

\subsection{Measurement equipment}
\label{sec:meas_equipment}

\begin{figure}[h]
    \centering     
    \plotFig{\plotFlag}{\resizebox{\columnwidth}{!}{\input{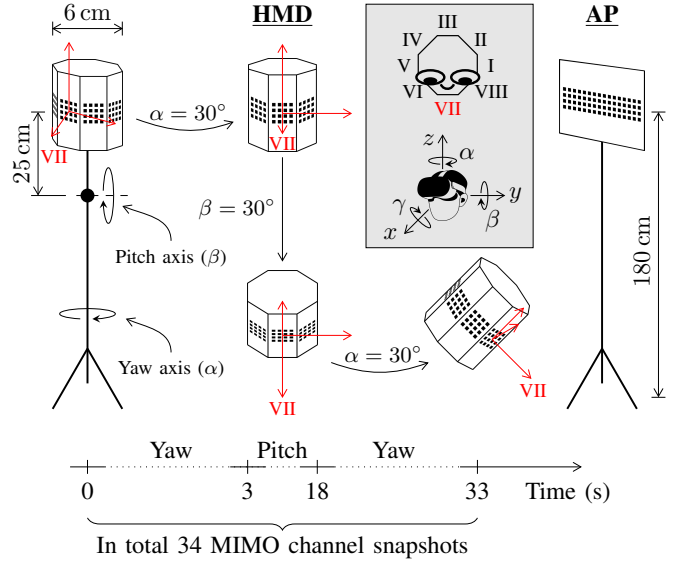}}}
    \caption{\gls{hmd} and \gls{ap} illustration, including the \glsposs{hmd} mobility pattern. \mbox{\gls{hmd} array} numbering and the extrinsic Euler rotation axes are shown in the gray rectangle. Positive rotations are obtained using the right-hand rule. Array VII points forwards (the same direction as the user's eyes and nose).}
    \label{fig:sounder_and_mobility_sequence}
    \vspacebelowfig
\end{figure}

\Cref{fig:sounder_and_mobility_sequence} shows the employed switched array \gls{mimo} channel sounder, jointly developed by Lund University and Sony. This sounder operates at \SI{28}{\giga\hertz}, has a bandwidth of \SI{768}{\mega\hertz} (\SI{1.3}{\nano\second} time\nobreakdash-delay resolution), and utilizes a Zadoff\nobreakdash-Chu waveform. The \gls{ue}, which we will refer to as the \gls{hmd}, consists of 8~planar patch\nobreakdash-antenna arrays, offset by \SI{45}{\degree} from each other in azimuth and featuring $4\times4$ antennas each. Throughout the measurement process, all 8~arrays are continuously sampled, and the channel matrices of specific arrays are extracted during the analysis stage in order to examine the channel performance of various \gls{hmd} array configurations. The \gls{hmd} arrays are marked by Roman numerals and VII represents the forward\nobreakdash-facing array, as shown in \Cref{fig:sounder_and_mobility_sequence}. The small array size (approximately $2\times2\,$\SI{}{\centi\meter}), makes it possible to mount eight arrays to an \glsposs{hmd} plastic frame or headband. Moreover, the arrays would halve in size for the \SI{60}{\giga\hertz} frequency band, and they would become even smaller in the sub\nobreakdash-THz spectrum. We foresee that arrays II and IV are the most critical, as these should not be placed behind the ears in order to avoid obstruction. The ears could otherwise constitute several \SI{}{\decibel} of attenuation, similar to how fingers obstruct the \gls{mmw} links on a smartphone \cite{xu_radiation_2018}. Dry thick hair is less of a problem with an attenuation of about \SI{0.2}{\decibel\per\centi\meter} (e.g., the cross\nobreakdash-section of a ponytail) \cite{corridon_does_2007}.

The \gls{ap}\nobreakdash-side is equipped with a single $16\times4$ planar array. All antenna elements on both the \gls{hmd} and \gls{ap} are dual\nobreak-polarized, resulting in a $256\times128$ channel matrix. Sampling is conducted at a rate of \SI{18.3}{\micro\second} per antenna element combination, which includes averaging over four sounding sequences to reduce measurement noise. The channel snapshot sampling rate is set to \SI{1}{\hertz}, providing sufficient time for sampling all antenna element pairs and for memory writing. Accurate synchronization is achieved by employing rubidium atomic clocks and additional preambles in the sounding waveform. More details on the channel sounder can be found in \cite{sounder_paper, cai_enhanced_2023}. The system parameters are listed in \Cref{tab:system_parameters}.

\begin{table}[h]
    \centering
    \vspace{-3mm}
    \caption{System parameters.}
    \vspace{-1mm}
    \begin{tabularx}{\columnwidth}{>{\hsize=.60\hsize}X>{\hsize=.40\hsize}Y}
    \toprule    
    \textbf{Parameter} & \textbf{Value} \\
    \midrule
    Carrier frequency & \SI{28}{\giga\hertz} \\
    Sounding signal bandwidth & \SI{768}{\mega\hertz}\\
    Time delay resolution & \SI{1.3}{\nano\second}\\
    Largest observable time delay & \SI{2.7}{\micro\second}\\
    Individual channel sampling time & \SI{18.3}{\micro\second} \\
    Channel snapshot sampling frequency & \SI{1}{\hertz} \\
    Measurement duration at each \gls{hmd} position & \SI{33}{\second} \\[0.75em]
    \gls{ap} array size & $4\times16$\\
    \gls{hmd} array size & $4\times4$\\
    Number of \gls{hmd} arrays & 8 \\ 
    Channel snapshot size 
    & $256\times128\times2048$ \\
    \bottomrule
\end{tabularx}
    \label{tab:system_parameters}
\end{table}

\subsection{Measurement environment}
\label{sec:meas_environment}

\begin{figure}[h]
     \centering
     \begin{subfigure}[b]{0.99\columnwidth}
        \centering              
        \plotFig{\plotFlag}{\resizebox{0.90\columnwidth}{!}{\input{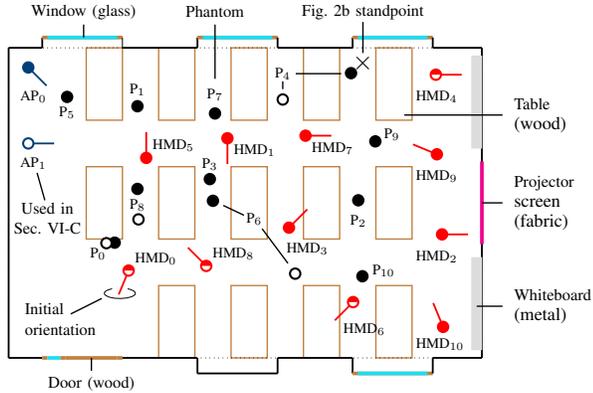}}}
        \vspace{-0.1 cm}
        \caption{Floor plan. Blue, red, and black markers represent the \gls{ap}, \gls{hmd}, and phantom, respectively. Filled and half-filled \gls{hmd} and phantom markers were considered for \gls{ap}$_0$, while empty (white) and \mbox{half-empty} markers were evaluated for \gls{ap}$_1$ (used in \Cref{sec:addconsres:impact_of_ap_position}). Brown, cyan, gray, and magenta mark tables and the door, windows, metal whiteboards, and a projector screen, respectively. The room size is $6\times9.15\;$\SI{}{\meter}.}
        \label{fig:measurement_environment_floorplan}
     \end{subfigure}
     \hfill
     \begin{subfigure}[b]{0.99\columnwidth}
        \centering     
        \plotFig{\plotFlag}{\resizebox{0.99\columnwidth}{!}{\input{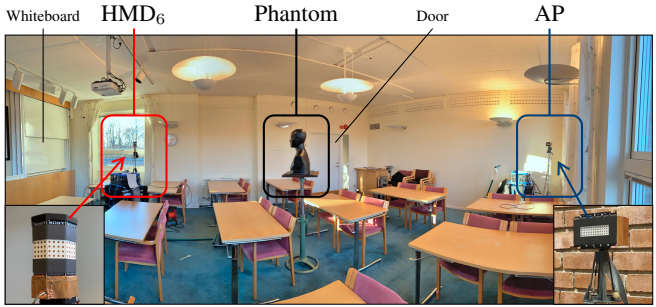}}}      
        \vspace{-0.2 cm}
        \caption{Setup for position \gls{hmd}$_6$ in \gls{olos} conditions.}
        \label{fig:measurement_environment_photo}
    \end{subfigure}
    \vspace{-0.6 cm}
    \caption{Conference room measurement environment.}
    \label{fig:measurement_environment}
    \vspacebelowfig
\end{figure}

Measurements were conducted at \SI{11}{}~fixed random positions in a conference room, as illustrated in \Cref{fig:measurement_environment}. For each position, the initial orientation of the \gls{hmd} was randomly chosen (indicated by the red line in \Cref{fig:measurement_environment_floorplan}). This initial orientation served as the starting point for the subsequent mobility pattern. Channel sounding was repeated at each position to simulate \gls{olos} conditions, where a pre\nobreakdash-characterised\footnote{Before the experiment, we evaluated the attenuation of the phantom as well as that of six volunteers by placing the \gls{ap} and \gls{hmd} \SI{3}{\meter} apart and positioning the obstructing subject in the center of the \gls{los}. The results from a laboratory environment showed a mean attenuation of \SI{12.8}{\decibel} for the phantom and $11.8\pm2.1$~\SI{}{\decibel} (mean and standard deviation) for the volunteers.} fiberglass water\nobreakdash-filled human phantom was used to obstruct the \gls{los} \cite{gustafson_characterization_2012}. The environment remained quasi\nobreakdash-static during channel sounding, except for the mobility of the \gls{hmd}, which will be discussed in the upcoming subsection.

\subsection{Mobility pattern}
\label{sec:mobility_pattern}

We have devised a mobility pattern for the \gls{hmd} that aims to capture \gls{mmw} channel information during directive antenna array rotation. As \Cref{fig:sounder_and_mobility_sequence} shows, the \gls{hmd} starts in an upright position and is first rotated by \SI{30}{\degree} in yaw (azimuth). 
This is followed by a \SI{30}{\degree} downward\nobreakdash-pitching movement (elevation, equivalent to leaning forward), and concluded by a final \SI{30}{\degree} yaw change. Rotation is executed manually using a tripod, to which the \gls{hmd} is mounted. The forward\nobreakdash-tilted \gls{hmd} is rotated around the tripod's vertical axis during the final yaw rotation. This mimics a person looking down and rotating their entire body by \SI{30}{\degree} in yaw (leftward rotation).

To ensure precise channel state information during rotation, we maintain angular velocity below \SI{10}{\degree\per\second} and tangential velocity below \SI{1}{\centi\meter\per\second}. Executing a single mobility sequence takes approximately \SI{33}{\second}. The first yaw rotation around the tripod's vertical axis takes \SI{3}{\second}, while the subsequent pitch and the final yaw rotation each take \SI{15}{\second}. The duration is longer for the second yaw rotation, since the \gls{hmd} covers a longer trajectory when it is tilted down, significantly increasing the radius of the movement. The \SI{25}{\centi\meter} distance between the pitch axis of the tripod and the \glsposs{hmd} center of mass closely reflects the separation between the human cervical vertebrae and the head \cite{kunin_rotation_2007}, making the mobility pattern a good representation of \gls{xr} user head movement. As noted in literature \cite{corbillon_360-degree_2017, fremerey_avtrack360_2018, blandino_head_2021}, roll rotation is far less pronounced than yaw and pitch rotation, therefore, we did not include roll in the mobility pattern.

\subsection{Signal model and MPC parameter estimation}
\label{sec:sage_estimation}

The \gls{ctf} for the $m^{\text{th}}$ \gls{hmd} and the $n^{\text{th}}$ \gls{ap} antenna is represented by a superposition of specular \glspl{mpc} (plane waves since the Fraunhofer distance is $<$\SI{1}{\meter}). The \gls{ctf} at the $k^{\text{th}}$ tone (frequency $f_k$) is modelled as
\begin{equation}
\begin{split}
\label{eq:channel_mpc_sum}
    & 
    \text{H}_{m,n}[p,s,i,k] \\ 
    &=  
    \sum_{l=0}^{L-1}\! 
    \text{\textbf{g}}_{\text{R},m}^\mathrm{T}(\phi_{\text{R},l}, 
    \theta_{\text{R},l}, f_k) 
    \begin{bmatrix}
    \gamma_{\text{HH},l} & \gamma_{\text{HV},l}\\
    \gamma_{\text{VH},l} & \gamma_{\text{VV},l}
    \end{bmatrix}
    \text{\textbf{g}}_{\text{T},n}(\phi_{\text{T},l}, 
    \theta_{\text{T},l}, f_k)
    \\
    &
    \times \text{g}(f_k) \,
    e^{-j 2\pi f_k \tau_l} \,
    e^{j 2\pi \nu_l t_{m,n}}
    + \mathcal{N}_{m,n}(f_k),
\end{split}
\end{equation}
where indexes $p$, $s$, $i$, and $k$ represent the \gls{hmd} position, scenario (\gls{los}/\gls{olos}), snapshot index, and sounding tone index (see \Cref{tab:index_overview} for a summary of indexes). Values $\text{\textbf{g}}_{\text{R},m} \! \in \! \mathbb{C} ^{2\times1}$ and $\text{\textbf{g}}_{\text{T},n} \! \in \! \mathbb{C} ^{2\times1}$ represent the frequency\nobreakdash-dependent polarimetric responses of the $m^{\text{th}}$ \gls{hmd} (Receiver) and the $n^{\text{th}}$ \gls{ap} (Transmitter) antenna, respectively. These further depend on the $l^{\text{th}}$ \glsposs{mpc} azimuth and elevation of arrival/departure, denoted by $\phi_{\text{R/T},l}$ and $\theta_{\text{R/T},l}$, correspondingly. Note that $(\cdot)^{\text{T}}$ is the matrix transpose. The parameters $\gamma_{\text{HH},l}$, $\gamma_{\text{HV},l}$, $\gamma_{\text{VH},l}$, and $\gamma_{\text{VV},l}$ represent the horizontal-to-horizontal, horizontal-to-vertical, vertical-to-horizontal, and vertical-to-vertical polarization amplitude gains of the $l^{\text{th}}$ \gls{mpc}. Furthermore, $g(f_k)$ represents the sounder's system response without antenna arrays, $\tau_l$ and $\nu_l$ are the $l^{\text{th}}$ \glsposs{mpc} time delay and Doppler frequency, respectively, $t_{m,n}$ is the sampling time, and $\mathcal{N}$ is white Gaussian noise. Note that $\text{H}_{m,n}$ is the Fourier transform of the \gls{cir}, represented by $\text{h}_{m,n}$.

\begin{table}[h]
    \centering
    \vspace{-3mm}
    \caption{Index overview.}
    \label{tab:index_overview}
    \vspace{-1mm}
    \begin{tabularx}{\columnwidth}{>{\hsize=.35\hsize}YX|>{\hsize=.35\hsize}YX}
    \toprule
    \textbf{Index} & \textbf{Meaning} & \textbf{Index} & \textbf{Meaning} \\
    \midrule
    p & \gls{hmd} position              & q & \gls{hmd} array \\
    s & \gls{los}/\gls{olos} Scenario   & m & \gls{hmd} antenna element \\
    i & Channel snapshot                & n & \gls{ap} antenna element \\
    k & Sounding tone             & j & Autocorrelation delay \\
    l & \gls{mpc} (signal component)    & r & Eigenmode (sp. stream) \\
    \bottomrule
\end{tabularx}
    \vspace{-4 mm}
\end{table}

We employ the \gls{sage} algorithm for estimating the channel's signal components and reconstructing a noiseless channel. While there is no reminiscent noise in the reconstructed channel, the estimation accuracy is subject to the ratio between channel gain after beamforming and the noise floor. This \gls{snr} is around \SI{50}{\decibel} across the measured channel snapshots. The corresponding error in the estimated channel parameters (Cramér\nobreakdash-Rao lower bound) is approximately inversely proportional to the aforementioned \gls{snr} after beamforming \cite{bellili_on_2010}. The reconstructed channel allows us to compare individual gains with higher accuracy and to later analyze channel capacity at different \glspl{snr}.

\subsection{Channel normalization}
\label{sec:meas_setp:channel_normalization}

For calculating the channel capacity, we apply channel normalization since removing large\nobreakdash-scale fading allows us to study capacity as a function of the \gls{snr}. This is done in order to solely assess the effects of limited \gls{fov} and the benefits of broadening the \gls{fov} by introducing additional antenna arrays. Normalized channel matrices are derived as follows 
\vspace{0.4 mm}
\begin{equation}
    \text{\textbf{H}}_{\scalebox{0.6}{$Q$}}[p,s,i,k] = \! \sqrt{ \frac{M\,N\,I\,K}{\sum\limits_{i=0}^{I-1}\sum\limits_{k=0}^{K-1}||\text{\textbf{H}}^{\text{rec}}_8[p,s,i,k]||_\mathrm{F}^2}} \text{\textbf{H}}^{\text{rec}}_{\scalebox{0.6}{$Q$}}[p,s,i,k],
\end{equation}
where $\text{\textbf{H}}_{\scalebox{0.6}{$Q$}}^{\text{rec}}$ is the reconstructed channel matrix for a $Q$\nobreakdash-array system without normalization and $||\cdot||_\mathrm{F}^2$ is the square of the Frobenius norm. Values $M=256$ and $N=128$ are the number of \gls{hmd} and \gls{ap} antenna ports, respectively, while $I=34$ and $K=2048$ represent the number of snapshots and sounding tones, correspondingly. The resulting mean channel gain over the frequency band becomes $E\,[\sum_{k=0}^{K-1}|\text{H}_{m,n}[p,s,i,k]|^2]=1$. We conduct normalization on the full $256\times128$~channel matrices (employing all 8~\gls{hmd} arrays), after which we extract the sub\nobreakdash-channels, corresponding to the studied \gls{hmd} antenna array configurations, in order to preserve their gain ratios for further evaluation. Note that this only applies to the spatial multiplexing capability and minimal service performance metrics, described in \Cref{sec:metrics:spatial_multiplexing} and \Cref{sec:metrics:minimal_service}, respectively. The preceding four gain\nobreakdash-based metrics evaluate the non\nobreakdash-normalized channel, i.e., $\text{\textbf{H}}^{\text{rec}}$.

Below is an example showing a full channel matrix and the sub\nobreakdash-channels of a single\nobreakdash-array configuration (discussed in \Cref{sec:analysis_procedures}). Indexes $p$, $s$, $i$, and $k$ are omitted for clarity.
\vspace{-1.4 mm}
\begin{equation}
\label{eq:channel_matrix_example}
\begin{tikzpicture}[baseline=(equation.center)]
\node (equation) at (0,0) {
    $\begin{aligned}
    \text{\textbf{H}} =
    \begin{bmatrix}
    \text{H}_{1,1} & \dots & \text{H}_{1,N}\\
    \vdots & \ddots & \vdots\\
    \text{H}_{193,1} & \dots & \text{H}_{193,N}\\
    \vdots & \ddots & \vdots\\
    \text{H}_{224,1} & \dots & \text{H}_{224,N}\\
    \vdots & \ddots & \vdots\\
    \text{H}_{M,1} & \dots & \text{H}_{M,N}\\
    \end{bmatrix}
    \end{aligned}$};
\draw [red, thick, rounded corners] ($(equation.west) + (+0.84,-0.8)$) rectangle ($(equation.east) + (-0.10,0.78)$);
\node [red, anchor=west, align=center] at ($(equation.east)$) {$\text{\textbf{H}}_1$\\Channel matrix of a \\single-array HMD\\(array VII)};
\end{tikzpicture}
\tag{3}
\end{equation}
\refstepcounter{equation}

\vspace{-8mm}\section{Performance metrics}
\label{sec:performacne_metrics}

This section describes the below six performance metrics. For each, we conduct an initial study based on analytical models, mathematically formulate the metric, and conclude with its application\nobreakdash-level relevance. 
Per sub\nobreakdash-section, the metrics are:
\begin{enumerate}[A.]
\vspace{0.2em}
\setlength\itemsep{0.5em}
    \item Gain dependency on the azimuth \gls{fov},
    \item Gain stability during rotation,
    \item Attenuation due to LoS obstruction,
    \item Delay dispersion and frequency-selective fading,
    \item Spatial multiplexing capability, and
    \item Minimal service.
\end{enumerate}

\subsection{Gain dependency on the azimuth FoV}
\label{sec:metrics:gain_tradeoff}

We first describe the \emph{power gain} (referred to as \emph{gain}) due to a signal component that impinges on an \gls{hmd}, equipped with multiple patch\nobreakdash-antenna arrays along its azimuth. To illustrate the relevance of the metric, we employ a theoretical analysis that assumes the antennas have no coupling between them and that they feature ideal responses with a \SI{-3}{\decibel}~\gls{fov} of~\SI{90}{\degree}. We express the compound gain as a superposition of the gains recorded by individual antennas on each of the arrays as
\vspace{-0.2mm} 
\begin{equation}
\label{eq:metrics:gain_tradeoff:limited_fov}
    \eta_{\scalebox{0.6}{$Q$}} (\alpha, \phi) = 
    N\!M_{\scalebox{0.6}{$Q$}}\!
    \sum\limits_{q=0}^{Q-1}
    |\alpha|^2
    \cos^2\!\!\!\;\left(\!\phi - q\frac{2\pi}{Q}\right)\!
    \left|\!\ceil*{\frac{|\phi - q\frac{2\pi}{Q}|}{\pi}}\!-2\right|,
\end{equation}
where $|\alpha|^2$ and $\phi$ are the gain and azimuth angle of the impinging signal component, while $Q$ represents the number of employed \gls{hmd} arrays, each consisting of $M_{\scalebox{0.6}{$Q$}}$ antennas. The impinging angle relative to each array's boresight is determined by $\phi - q\frac{2\pi}{Q}$, while the last term in \Cref{eq:metrics:gain_tradeoff:limited_fov} limits the radiation pattern to $\pm\frac{\pi}{2}$. We see that reducing the number of arrays ($Q$) does not necessarily decrease the compound gain by a factor of $Q$ if the channel's \glspl{mpc} are unequally distributed along the azimuth. For example, a single\nobreakdash-array configuration might receive similar gain to a 4\nobreakdash-array configuration when oriented towards the \gls{los}. However, orienting the single\nobreakdash-array configuration's narrow \gls{fov} away from the channel's major \glspl{mpc} would result in a far lower gain than suggested merely by the reduction in array gain. To evaluate the effects of a limited compound \gls{hmd} \gls{fov} on the gain for the empirical data, we derive the gain ratio between a $Q$\nobreakdash-array and a full 8\nobreakdash-array configuration. Considering all \glspl{mpc} per channel snapshots, we drop parameters $\alpha$ and $\phi$ to derive the ratio as
\vspace{-0.1mm} 
\begin{equation}
\label{eq:metrics:gain_tradeoff}
    \overline{\Delta\eta_{\scalebox{0.6}{$Q$}}}[p,s,i] =  \frac{\overline{\eta_{\scalebox{0.6}{$Q$}}}[p,s,i]}{\overline{\eta_{8}}[p,s,i]},
\vspace{-0.2mm} 
\end{equation}
where $\overline{\eta_{\scalebox{0.6}{$Q$}}}[p,s,i]$ is the mean power gain recorded by a $Q$\nobreakdash-array \gls{hmd} across the \SI{768}{\mega\hertz} bandwidth. If the $Q$\nobreakdash-array configuration is illuminated by the \gls{los}, then the metric yields $\overline{\Delta\eta_{\scalebox{0.6}{$Q$}}} \geq \frac{\scalebox{0.6}{$Q$}}{8}$ since only a few of the eight arrays of the full configuration can capture the \gls{los} simultaneously. The result becomes $\overline{\Delta\eta_{\scalebox{0.6}{$Q$}}} < \frac{\scalebox{0.6}{$Q$}}{8}$ if the $Q$\nobreakdash-array configuration's \gls{fov} can not capture the major \glspl{mpc}, including the \gls{los}.

\textit{Application-level relevance:} An \gls{fov} which is too narrow leads to a pronounced reduction in gain. The resulting data rate decrease can ripple through the application layer, reducing \gls{xr} video quality and possibly causing outage.

\subsection{Gain stability during rotation}
\label{sec:metrics:gain_stability}

The time dynamics of channel gain during \gls{hmd} rotation are assessed through gain spread and persistence. We derive these, respectively, by calculating the standard deviation and the autocorrelation of the gain across the \gls{hmd} orientations (channel snapshots) that make up the mobility pattern. The gain's standard deviation is derived in the logarithmic domain (in decibels) in order to better highlight the relative gain changes, regardless of the mean gain, according to the below
\begin{equation}
\label{eq:metrics:gain_stability:standard_deviation}
    \sigma_{\eta, {\scalebox{0.6}{$Q$}}}[p,s] = \frac{10}{I} \sqrt{  \, \sum\limits_{i=0}^{I-1} \log^{2}_{10} \left( \cfrac{ \overline{\eta_{\scalebox{0.6}{$Q$}}}[p,s,i]}{\overline{\overline{\eta_{\scalebox{0.6}{$Q$}}}}[p,s]} \right)} \quad\quad [\mathrm{dB}],
\end{equation}
where $\overline{\overline{\eta_{\scalebox{0.6}{$Q$}}}}[p,s] = \mathbb{E}[\:\!\overline{\eta_{\scalebox{0.6}{$Q$}}}[p,s,i]]$ is the mean gain across all snapshots for a given position\nobreakdash-scenario (\gls{los}/\gls{olos}) pair. We calculate the gain's autocorrelation as a normalized value in the range of $r \in (-1,1)$ as follows
\begin{equation}
\begin{gathered}
\label{eq:metrics:gain_stability:autocorrelation}
    r_{\eta, {\scalebox{0.6}{$Q$}}}[p,s,j] = \\ \frac
    { \sum\limits_{i=0}^{I-j-1} ( \overline{\eta_{\scalebox{0.6}{$Q$}}}[p,s,i]\!-\!\overline{\overline{\eta_{\scalebox{0.6}{$Q$}}}}[p,s] ) ( \overline{\eta_{\scalebox{0.6}{$Q$}}}[p,s,i\!+\!j]\!-\!\overline{\overline{\eta_{\scalebox{0.6}{$Q$}}}}[p,s] ) }
    { \sum\limits_{i=j}^{I-j-1} ( \overline{\eta_{\scalebox{0.6}{$Q$}}}[p,s,i] - \overline{\overline{\eta_{\scalebox{0.6}{$Q$}}}}[p,s] )^2 },
\end{gathered}
\end{equation}
where $j$ represents the delay in terms of channel snapshots.  

\textit{Application-level relevance:} A high standard deviation of the gain during the conducted rotation may lead to video resolution adaptation. Moreover, a low autocorrelation shows low gain stability (frequent changes), which may require additional video frame buffering. This leads to higher \gls{xr} video latency.

\subsection{Attenuation due to LoS obstruction}
\label{sec:metrics:gain_attenuation} 
\Gls{los} obstruction by a person (or phantom) other than the \gls{hmd} user causes substantial attenuation. We anticipate the latter will conform with the \gls{gtd}\footnote{A detailed description of the \gls{gtd} can be found in \cite{gustafson_characterization_2012} and \cite{james_geometrical_1986}.}, which describes attenuation according to the obstructing object's shape and its distance to the \gls{ap} and \gls{hmd} \cite{gustafson_characterization_2012}. The discrepancy from the \gls{gtd}, presented alongside the results, will highlight the array configuration's capability of capturing additional unobstructed \glspl{mpc}. To determine the attenuation upon \gls{los} obstruction, we compare the \gls{los} channel gain against that of the \gls{olos} scenario using
\begin{equation}
\label{eq:obstruction}
    \overline{\Delta\eta_{\scalebox{0.6}{$Q$}}^{\text{(bl)}}}[p, I] = \frac
    {\sum\limits_{i=0}^{I-1}\overline{\eta_{\scalebox{0.6}{$Q$}}}[p,\text{LoS},i]}
    {\sum\limits_{i=0}^{I-1}\overline{\eta_{\scalebox{0.6}{$Q$}}}[p,\text{OLoS},i]},
\end{equation}
where the total number of evaluated snapshots $I$ is in this assessment limited to either one or all available snapshots. We first evaluate for $I=1$ which gives us a comparison of the channel gain of the first snapshot at each position. This is done since at the start of each measurement, the \gls{hmd} part of the channel sounder is approximately in the same position and orientation for both the \gls{los} and \gls{olos} scenario. The orientation of the \gls{hmd} can vary during subsequent snapshots due to the manual execution of the mobility sequence. In the second step of the assessment, $I$ is increased to incorporate all available snapshots per position and scenario. This results in a comparison of the mean gain during \gls{los} and \gls{olos} conditions. 

\textit{Application-level relevance:} \Gls{los} obstruction by another person's head or body can cause substantial attenuation. This metric shows how severe the attenuation is and whether the \gls{hmd} can benefit from receiving other unobstructed \glspl{mpc}.

\subsection{RMS delay spread and frequency-selective fading}
\label{sec:metrics:delay_spread}

A reflective environment can lead to the transmitted signal arriving at the receiver at multiple instances in time due to multipath propagation. Its extent can be summarized using the \gls{rms} delay spread, calculated as follows 
\begin{equation}
\begin{gathered}
\label{eq:rms_delay_spread}
    \sigma_{\tau,q}[p,s,i] = \\[4pt]
    \sqrt{ \frac{ \sum\limits_{u=0}^{U-1}||\text{\textbf{h}}_{\scalebox{0.6}{$Q$}}[u]||_\mathrm{F}^2\tau[u]^2 }{ \sum\limits_{u=0}^{U-1}||\text{\textbf{h}}_{\scalebox{0.6}{$Q$}}[u]||_\mathrm{F}^2} - \left( \frac{ \sum\limits_{u=0}^{U-1}||\text{\textbf{h}}_{\scalebox{0.6}{$Q$}}[u]||_\mathrm{F}^2\tau[u] }{ \sum\limits_{u=0}^{U-1}||\text{\textbf{h}}_{\scalebox{0.6}{$Q$}}[u]||_\mathrm{F}^2} \right)^{\!\!2} },
\end{gathered}
\end{equation}
where $||\text{\textbf{h}}_{\scalebox{0.6}{$Q$}}[u]||_\mathrm{F}^2$ is the \gls{pdp} of a $Q$\nobreakdash-array \gls{hmd} and $\tau[u]$ is the time delay of the $u^{\text{th}}$ \gls{cir} sample. The corresponding coherence bandwidth is approximately inversely proportional to the channel's \gls{rms} delay spread \cite{molisch_wireless_book}. When the transmission bandwidth is wider than the coherence bandwidth, frequency\nobreakdash-selective fading occurs. Its severity is assessed through the gain's persistence over the measured bandwidth, calculated as follows
\begin{equation}
\label{eq:frequency_sel_fad}
    \xi_{\scalebox{0.6}{$Q$}}[p,s,i] = \frac{\sum\limits_{k=0}^{K-1}\left(\eta_{\scalebox{0.6}{$Q$}}[p,s,i,k] - \overline{\eta_{\scalebox{0.6}{$Q$}}}[p,s,i]\right)^2}{\overline{\eta_{\scalebox{0.6}{$Q$}}}[p,s,i]^2}.
\end{equation}
Recall that $\eta_{\scalebox{0.6}{$Q$}}$ is the gain of the compound \gls{mimo} channel across multiple \gls{ap}\nobreakdash-\gls{hmd} antenna pairs. The intuition behind the metric is that the ratio between the gain's standard deviation and its mean across the transmission bandwidth reduces as more channels are included. For ideally persistent gain across
the measured bandwidth (flat fading), $\xi$ would reduce to zero.

\textit{Application-level relevance:} The \gls{rms} delay spread indicates the coherence bandwidth within which roughly flat fading is expected. This highlights the needed complexity of the channel equalization with regards to the employed transmission bandwidth. The latter depends on the required data rate.

\subsection{Spatial multiplexing capability}
\label{sec:metrics:spatial_multiplexing}

Transmission can take place over a single stream, called \gls{det}, or over several streams, referred to as \gls{sm}. The maximal number of multiplexed streams is governed by the channel as $R\leq\min(M,N)$. \Gls{sm} exploits the channel's rank (orthogonality) to increase channel capacity as follows 
\begin{equation}
\label{eq:minimal_service}
    C_{\scalebox{0.6}{$Q$}}[p,s,i] = \max_{\sum \rho_{\scalebox{0.6}{$Q$},r} \leq M_{\scalebox{0.6}{$Q$}}} \sum_{r=0}^{R_{\scalebox{0.6}{$Q$}}-1} \log_2 \left( 1 + \frac{E_s\,\rho_{\scalebox{0.6}{$Q$},r}}{M_{\scalebox{0.6}{$Q$}}\,N_0} \overline{\lambda_{\scalebox{0.6}{$Q$},r}} \right),
\end{equation}
where $R_{\scalebox{0.6}{$Q$}}$ is the rank of channel matrix $\text{\textbf{H}}_{\scalebox{0.6}{$Q$}}$ (number of non\nobreakdash-zero eigenmodes), $\overline{\lambda_{\scalebox{0.6}{$Q$},r}}$ is the mean $r^{\text{th}}$ eigenvalue of $\text{\textbf{H}}_{\scalebox{0.6}{$Q$}}\text{\textbf{H}}_{\scalebox{0.6}{$Q$}}^{\text{T}}$, $M_{\scalebox{0.6}{$Q$}}$ is the number of channel outputs (\gls{hmd} antenna ports), $\frac{E_s}{N_0}$ is the symbol energy over noise spectral density, and the power allocated to the $r^{\text{th}}$ eigenmode (orthogonal stream) is represented by $\rho_{\scalebox{0.6}{$Q$},r}$ and determined according to the waterfilling algorithm \cite{molisch_wireless_book}. Capacity is largest for a full\nobreakdash-rank channel with equal\nobreakdash-gain eigenmodes (unitary channel). A \gls{los} channel may also have a full\nobreakdash-rank, however, its gain is often severely unevenly distributed among its eigenmodes \cite{marzetta_fundamentals_2016}. Maximizing capacity according to waterfilling leads us to allocate power to a limited number of streams in such a channel, whereas power would be equally distributed among the eigenmodes of a unitary channel. To evaluate the \gls{sm} capabilities of the considered array configurations and to assess the profitability of allocating power to more than one stream, we progressively increment the number of employed streams ($R_{\scalebox{0.6}{$Q$}}$) in \Cref{eq:minimal_service} and observe the obtained capacity. Note that $R_{\scalebox{0.6}{$Q$}}=1$ reduces \Cref{eq:minimal_service} to \gls{det}, equivalent to joint maximum\nobreakdash-ratio transmission and combining (MRT/MRC).

\textit{Application-level relevance:} Multiplexing several streams can increase the data rate and enhance \gls{xr} video quality. Knowing the increase in capacity per additional stream allows \gls{hmd} designers to appropriately scale the system's complexity and avoid overprovisioning of the hardware.

\subsection{Minimal service}
\label{sec:metrics:minimal_service}

The channel capacity derived using \Cref{eq:minimal_service} also serves as the upper bound of the achievable data rate. Typically, channel characterization studies evaluate the mean and sum (for multi\nobreakdash-user scenarios) channel capacity \cite{flordelis_spatial_2018}. However, we foresee that for \gls{xr} applications, where video stream reception represents the main payload, it is more important to study the minimal capacity, since it determines the worst\nobreakdash-case video quality. As most use cases, \gls{xr} applications allow for a limited amount of outage. Namely, video streaming has an approximately \SI{97}{\percent} reliability requirement, allowing up to \SI{3}{\percent} of outage \cite{nightingale_subjective_2013}. We thus primarily focus on analyzing the capacity's 3$^{\text{rd}}$ percentile and on evaluating how this value is affected by the number of employed antenna arrays. Moreover, considering channel gain normalization, we study the minimal service's dependency on the receive \gls{snr} by varying $\frac{E_s}{N_0}$.

\Cref{eq:minimal_service} employs mean eigenmode gain, averaged across the transmission bandwidth. This is a measure of the potential achievable rate over the full transmission bandwidth. However, individual subcarriers might undergo more pronounced fading. We assess both approaches in order to evaluate whether the same trends in capacity occur when increasing the number of employed arrays. We derive the minimal service on a per\nobreakdash-subcarrier basis (\SI{375}{\kilo\hertz} spacing) by evaluating the statistics of $C_{\scalebox{0.6}{$Q$}}[p,s,i,k]$, without eigenmode averaging.

\textit{Application-level relevance:} Channel capacity provides the upper bounds of the data rate, which in terms limits the achievable \gls{xr} video quality. Since video streaming has a \SI{97}{\percent} reliability constraint, the capacity's 3$^{\text{rd}}$ percentile determines the minimal service, that is, the minimal relevant capacity \cite{nightingale_subjective_2013}.

\section{Analysis procedures}
\label{sec:analysis_procedures}
This section introduces the analysis procedures, utilizing the performance metrics outlined in the preceding section and applying them to channel measurement data. The following sections describe our approach to:

\begin{enumerate}[A.]
\vspace{0.2em}
\setlength\itemsep{0.5em}
    \item Evaluating the performance of a multi\nobreakdash-array \gls{hmd},
    \item Investigating the profitability of a rear\nobreakdash-\gls{hmd} headband,
    \item Assessing performance degradation when activating a limited number of antennas in \gls{hmd} arrays, and
    \item Exploring the impact of the \glsposs{ap} position on the channel.
\end{enumerate} 
\vspace{-0.2em}

\subsection{Multi-array HMD performance}
\label{sec:aproc:mutli_array_benefits}

\begin{figure}[h]
    \centering    
    \plotFig{\plotFlag}{\resizebox{\columnwidth}{!}{\input{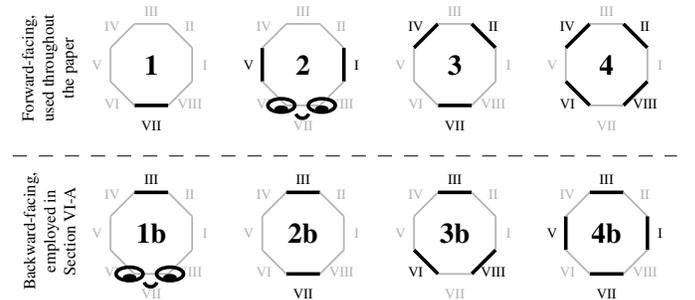}}}  
    \caption{Studied \gls{hmd} array configurations, top-down view. The top and bottom row show the forward- and backward-facing configurations, respectively.}
    \label{fig:array_configurations_fwd}
    \vspacebelowfig
\end{figure}

The primary focus of the study at hand is to estimate the performance of an \gls{hmd}, which, in its final design, would feature between 1 and 4 \gls{mmw} antenna arrays. This is achieved by applying the proposed performance metrics from \Cref{sec:performacne_metrics} to \gls{mmw} channel measurements for the antenna array configurations illustrated in \Cref{fig:array_configurations_fwd}. The four configurations are primarily forward\nobreakdash-oriented, as we envision that future \glspl{hmd} will not feature a rear headband in order to prioritize ergonomics and aesthetics. As concluded in our initial analysis \cite{marinsek_impact_2023}, there is no clear benefit to including more than \SI{4}{} antenna arrays, apart from array gain. This is supported by the fact that \SI{4}{}~patch\nobreakdash-antenna arrays already provide approximately uniform \SI{360}{}\nobreakdash-degree coverage.

\subsection{Utilizing a rear headband}
\label{sec:aproc:read_headband}

A considerable number of present\nobreakdash-day \glspl{hmd} incorporate a rear headband. Although this makes the \gls{hmd} more conspicuous, it does offer additional mounting space for antenna arrays. We evaluate the benefits of featuring a rear \gls{hmd} headband and utilizing it for array placement by comparing the forward\nobreakdash-facing configurations in \Cref{fig:array_configurations_fwd} with their backward\nobreakdash-facing counterparts. 
The profitability of the rear headband is established by evaluating the gain ratio as follows
\begin{equation}
    \overline{\Delta\eta_{\scalebox{0.6}{$Q$}}^{\text{(rh)}}}[p,s,i] =  \frac{\overline{\eta_{\scalebox{0.6}{$Q$}}^{\text{(b)}}}[p,s,i]}{\overline{\eta_{\scalebox{0.6}{$Q$}}^{\text{(f)}}}[p,s,i]},
\end{equation}
where $\overline{\eta_{\scalebox{0.6}{$Q$}}^{\text{(f)}}}$ and $\overline{\eta_{\scalebox{0.6}{$Q$}}^{\text{(b)}}}$ represent the mean gain of a forward- or backward\nobreakdash-facing $Q$\nobreakdash-array configuration, respectively.

\subsection{Activating a limited number of antennas in HMD arrays}
\label{sec:aproc:deactivating_antennas}

\begin{figure}[h]
    \centering
    \vspace{-3mm}
    \plotFig{\plotFlag}{\resizebox{\columnwidth}{!}{\begin{tikzpicture}[every node/.append style={font=\fontsize{11}{14}\selectfont}]


\definecolor{blue1}{RGB}{0,64,122}
\definecolor{blue2}{RGB}{82,189,236}
\definecolor{red1}{RGB}{196, 30, 58}
\definecolor{red2}{RGB}{238, 75, 43}
\definecolor{red3}{RGB}{220, 20, 60}

\definecolor{gray1}{RGB}{180,180,180}
\definecolor{gray2}{RGB}{140,140,140}
\definecolor{cyan1}{RGB}{27, 221, 242}
\definecolor{orange1}{RGB}{221,138,46}
\definecolor{orange2}{RGB}{232,179,3}
\definecolor{brown1}{RGB}{84,40,17}
\definecolor{copper}{RGB}{184, 115, 51}
\definecolor{gold}{RGB}{214, 185, 114}

\def\RoomLength{9.15}
\def\RoomWidth{6.00}
\def\SideWbWidth{1.80}
\def\WindowWidth{1.35}

\def\TableWidth{1.40}
\def\TableDepth{.70}
\def\TableSpacing{.70}
\def\TableOffset{1.50}
\def\EisleWidth{.90}

\def\OrientationOffset{90} 

\def\NodeSize{0.12}

\def\antennaSize{.2}
\def\arraySize{\antennaSize*6.5}
\def\arrayRegion{\antennaSize*8}
\newcommand\Square[1]{+(-#1,-#1) rectangle +(#1,#1)}


\def\xoffset{1.20*0*\arrayRegion}
\def\yoffset{0*\arrayRegion}
\def\y{2}
\foreach \x in {1,...,1}{
    \draw[fill=gold] (\xoffset+\x*\antennaSize*1.5,\yoffset+\y*\antennaSize*1.5) rectangle ++ (\antennaSize, \antennaSize);
}
\draw[draw=black, line width=0.75] (\xoffset-\antennaSize*0.5,\yoffset-\antennaSize*0.5) rectangle ++ (\antennaSize*6.5, \antennaSize*6.5);
\foreach \y in {0,...,3}{
    \foreach \x in {0,...,3}{
        \draw[draw=black, line width=0.50] (\xoffset+\x*\antennaSize*1.5,\yoffset+\y*\antennaSize*1.5) rectangle ++ (\antennaSize, \antennaSize);
    }
}

\def\xoffset{1.20*1*\arrayRegion}
\def\yoffset{0.48*\arrayRegion}
\def\y{2}
\foreach \x in {1,...,2}{
    \draw[fill=gold] (\xoffset+\x*\antennaSize*1.5,\yoffset+\y*\antennaSize*1.5) rectangle ++ (\antennaSize, \antennaSize);
}
\draw[draw=black, line width=0.75] (\xoffset-\antennaSize*0.5,\yoffset-\antennaSize*0.5) rectangle ++ (\antennaSize*6.5, \antennaSize*6.5);
\foreach \y in {0,...,3}{
    \foreach \x in {0,...,3}{
        \draw[draw=black, line width=0.50] (\xoffset+\x*\antennaSize*1.5,\yoffset+\y*\antennaSize*1.5) rectangle ++ (\antennaSize, \antennaSize);
    }
}

\def\xoffset{1.20*1*\arrayRegion}
\def\yoffset{-0.48*\arrayRegion}
\def\x{1}
\foreach \y in {1,...,2}{
    \draw[fill=gold] (\xoffset+\x*\antennaSize*1.5,\yoffset+\y*\antennaSize*1.5) rectangle ++ (\antennaSize, \antennaSize);
}
\draw[draw=black, line width=0.75] (\xoffset-\antennaSize*0.5,\yoffset-\antennaSize*0.5) rectangle ++ (\antennaSize*6.5, \antennaSize*6.5);
\foreach \y in {0,...,3}{
    \foreach \x in {0,...,3}{
        \draw[draw=black, line width=0.50] (\xoffset+\x*\antennaSize*1.5,\yoffset+\y*\antennaSize*1.5) rectangle ++ (\antennaSize, \antennaSize);
    }
}

\def\xoffset{1.20*2*\arrayRegion}
\def\yoffset{-0.98*\arrayRegion}
\def\y{2}
\foreach \x in {0,...,3}{
    \draw[fill=gold] (\xoffset+\x*\antennaSize*1.5,\yoffset+\y*\antennaSize*1.5) rectangle ++ (\antennaSize, \antennaSize);
}
\draw[draw=black, line width=0.75] (\xoffset-\antennaSize*0.5,\yoffset-\antennaSize*0.5) rectangle ++ (\antennaSize*6.5, \antennaSize*6.5);
\foreach \y in {0,...,3}{
    \foreach \x in {0,...,3}{
        \draw[draw=black, line width=0.50] (\xoffset+\x*\antennaSize*1.5,\yoffset+\y*\antennaSize*1.5) rectangle ++ (\antennaSize, \antennaSize);
    }
}

\def\xoffset{1.20*2*\arrayRegion}
\def\yoffset{0.98*\arrayRegion}
\def\x{1}
\foreach \y in {0,...,3}{
    \draw[fill=gold] (\xoffset+\x*\antennaSize*1.5,\yoffset+\y*\antennaSize*1.5) rectangle ++ (\antennaSize, \antennaSize);
}
\draw[draw=black, line width=0.75] (\xoffset-\antennaSize*0.5,\yoffset-\antennaSize*0.5) rectangle ++ (\antennaSize*6.5, \antennaSize*6.5);
\foreach \y in {0,...,3}{
    \foreach \x in {0,...,3}{
        \draw[draw=black, line width=0.50] (\xoffset+\x*\antennaSize*1.5,\yoffset+\y*\antennaSize*1.5) rectangle ++ (\antennaSize, \antennaSize);
    }
}

\def\xoffset{1.20*2*\arrayRegion}
\def\yoffset{0*\arrayRegion}
\foreach \y in {1,...,2}{
    \foreach \x in {1,...,2}{
        \draw[fill=gold] (\xoffset+\x*\antennaSize*1.5,\yoffset+\y*\antennaSize*1.5) rectangle ++ (\antennaSize, \antennaSize);
    }
}
\draw[draw=black, line width=0.75] (\xoffset-\antennaSize*0.5,\yoffset-\antennaSize*0.5) rectangle ++ (\antennaSize*6.5, \antennaSize*6.5);
\foreach \y in {0,...,3}{
    \foreach \x in {0,...,3}{
        \draw[draw=black, line width=0.50] (\xoffset+\x*\antennaSize*1.5,\yoffset+\y*\antennaSize*1.5) rectangle ++ (\antennaSize, \antennaSize);
    }
}

\def\xoffset{1.20*3*\arrayRegion}
\def\yoffset{0.48*\arrayRegion}
\foreach \y in {1,...,2}{
    \foreach \x in {0,...,3}{
        \draw[fill=gold] (\xoffset+\x*\antennaSize*1.5,\yoffset+\y*\antennaSize*1.5) rectangle ++ (\antennaSize, \antennaSize);
    }
}
\draw[draw=black, line width=0.75] (\xoffset-\antennaSize*0.5,\yoffset-\antennaSize*0.5) rectangle ++ (\antennaSize*6.5, \antennaSize*6.5);
\foreach \y in {0,...,3}{
    \foreach \x in {0,...,3}{
        \draw[draw=black, line width=0.50] (\xoffset+\x*\antennaSize*1.5,\yoffset+\y*\antennaSize*1.5) rectangle ++ (\antennaSize, \antennaSize);
    }
}

\def\xoffset{1.20*3*\arrayRegion}
\def\yoffset{-0.48*\arrayRegion}
\foreach \y in {0,...,3}{
    \foreach \x in {1,...,2}{
        \draw[fill=gold] (\xoffset+\x*\antennaSize*1.5,\yoffset+\y*\antennaSize*1.5) rectangle ++ (\antennaSize, \antennaSize);
    }
}
\draw[draw=black, line width=0.75] (\xoffset-\antennaSize*0.5,\yoffset-\antennaSize*0.5) rectangle ++ (\antennaSize*6.5, \antennaSize*6.5);
\foreach \y in {0,...,3}{
    \foreach \x in {0,...,3}{
        \draw[draw=black, line width=0.50] (\xoffset+\x*\antennaSize*1.5,\yoffset+\y*\antennaSize*1.5) rectangle ++ (\antennaSize, \antennaSize);
    }
}

\def\xoffset{1.20*4*\arrayRegion}
\def\yoffset{0*\arrayRegion}
\foreach \y in {0,...,3}{
    \foreach \x in {0,...,3}{
        \draw[fill=gold] (\xoffset+\x*\antennaSize*1.5,\y*\antennaSize*1.5) rectangle ++ (\antennaSize, \antennaSize);
    }
}
\draw[draw=black, line width=0.75] (\xoffset-\antennaSize*0.5,\yoffset-\antennaSize*0.5) rectangle ++ (\antennaSize*6.5, \antennaSize*6.5);
\foreach \y in {0,...,3}{
    \foreach \x in {0,...,3}{
        \draw[draw=black, line width=0.50] (\xoffset+\x*\antennaSize*1.5,\yoffset+\y*\antennaSize*1.5) rectangle ++ (\antennaSize, \antennaSize);
    }
}

\def\yoffset{-1.3*\arrayRegion}

\node[anchor=center, align=center, font=\footnotesize] at (1.20*0*\arrayRegion-\antennaSize*0.5+\arraySize*0.5, 0.9*\arrayRegion) {1 antenna/array};
\node[anchor=center, align=center, font=\footnotesize] at (1.20*1*\arrayRegion-\antennaSize*0.5+\arraySize*0.5, 1.4*\arrayRegion) {2 antennas/array};
\node[anchor=center, align=center, font=\footnotesize] at (1.20*2*\arrayRegion-\antennaSize*0.5+\arraySize*0.5, 1.9*\arrayRegion) {4 antennas/array};
\node[anchor=center, align=center, font=\footnotesize] at (1.20*3*\arrayRegion-\antennaSize*0.5+\arraySize*0.5, 1.4*\arrayRegion) {8 antennas/array};
\node[anchor=center, align=center, font=\footnotesize] at (1.20*4*\arrayRegion-\antennaSize*0.5+\arraySize*0.5, 0.9*\arrayRegion) {16 antennas/array};

\end{tikzpicture}}} 
    \caption{Examples of the evaluated rectangular \gls{hmd} antenna sub-array shapes (active antennas are marked yellow). For each sub-array, all possible locations within the full $4\times4$ array are included in the analysis.}
    \label{fig:varying_hmd_array_size}
    \vspacebelowfig
\end{figure}
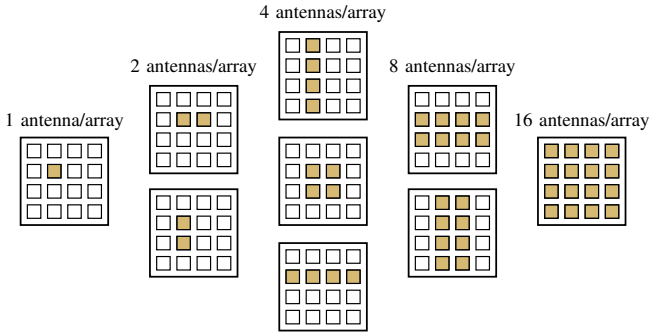

Wireless \glspl{hmd} are energy\nobreakdash-constrained devices, and powering off the \gls{rf} chains of individual antennas can reduce their energy consumption. 
The drawback is a possibly lower spatial diversity gain due to the smaller number of combined signals, which may simultaneously suffer from fading  \cite{morais_5g_2023}. 
To evaluate the impact of powering off \gls{rf} chains, we analyze the channel in terms of the six performance metrics for an \gls{hmd} that utilizes 1\nobreakdash--16 antennas per array, as shown in \Cref{fig:varying_hmd_array_size}. We conduct the analysis by selecting the same antennas on all \gls{hmd} arrays (e.g., the same row on all $Q$\nobreakdash-arrays for $1\times4$) and evaluating their corresponding channel matrix. We analyze only the 3\nobreakdash-array configuration, which showed prosperous performance when using all 16~antennas per array, as reported later in \Cref{sec:res_antenna_array_configuration}. We repeat the evaluation for all possible sub\nobreakdash-array positions and report the joint results. For instance, we examine four different $1\times4$ sub\nobreakdash-arrays, corresponding to the four rows in the $4\times4$ array. The results in \Cref{sec:addconsres:deactivating_antennas} show the joint statistical distributions. We employ rectangular sub\nobreakdash-arrays with equidistant antenna spacing in order to also assess the differences between vertical and horizontal shapes.

In addition to the six performance metrics, we assess the dependency of channel gain correlation on the number of active \gls{hmd} antennas, providing a measure of how likely the channels are to simultaneously suffer from fading. First, we derive the channel gain covariance matrix as per~\cite{paulraj_introduciton_2003} as
\begin{equation}
\begin{gathered}
\label{eq:aproc:deactivating_antennas:covariance}
    \mathbf{R}[p,s,i] = \\[-0.4em]
    \frac{1}{K} \sum\limits_{k=0}^{K-1}
    |\mathrm{vec}(\hat{\mathbf{H}}_{\mathrm{rec}}[p,s,i,k])|^2 \, 
    |\mathrm{vec}(\hat{\mathbf{H}}_{\mathrm{rec}}[p,s,i,k])^{\mathrm{H}}|^2,
\end{gathered}
\end{equation}
where $\mathrm{vec}(\hat{\mathbf{H}}_{\mathrm{rec}}[p,s,i,k])$ is the recorded sub\nobreakdash-channel matrix at the $k^{\text{th}}$ sounding tone, reorganized from an \mbox{$\hat{M} \times \hat{N}$} matrix into an $\hat{M}\hat{N} \times 1$ vector. \Cref{eq:aproc:deactivating_antennas:covariance} thus derives the gain covariance among channels by multiplying the channel gains of each combination of two channels at every sounding tone. We normalize the covariance matrix and calculate its mean to obtain the mean channel gain correlation per channel snapshot. The decrease in correlation, observed with the use of additional antennas (from 1 to 16), shows the dependency of spatial diversity on the number of active \gls{rf} chains.

\subsection{Impact of AP position and path loss assessment}
\label{sec:aproc:ap_position}

\begin{table*}[t]
    \centering
    \renewcommand{\arraystretch}{1.5}
    \caption{Overview of the employed performance metrics and the main insight for \gls{mmw} \glspl{hmd}. The column \textit{Sec.} lists the sections that contain the results corresponding to the described performance metrics.}
    \label{tab:performance_metric_overview}
    \begin{tabular}{>{\raggedright}p{2.7cm}|c|p{7cm}|p{6.4cm}}
    \toprule
    \textbf{Metric} & \textbf{Sec.} & \textbf{Description} & \textbf{Insight for \gls{mmw} \glspl{hmd}} \\
    \midrule
    
    \rowcolor{wr_gray} Gain dependency on \mbox{the azimuth field of view} & \ref{sec:configres:gain_tradeoff} & Reduction in gain, relative to a full \SI{360}{\degree} azimuth \gls{fov}. Low values indicate that the configuration can not capture the dominant \glspl{mpc}. & A single planar \gls{hmd} array can not capture sufficient \glspl{mpc}. Multiple \gls{hmd} arrays or distributed \glspl{ap} are required.\\
    
    \rowcolor{pw_gray} Gain stability \mbox{during rotation} & \ref{sec:configres:gain_stability} & The gain's standard deviation and autocorrelation during rotation. Small std. deviation and high autocorrelation indicate good stability. & Already two arrays reduce gain fluctuation, while adding more arrays further decreases gain spread with diminishing returns. \\
    
    \rowcolor{wr_gray} Attenuation due to \mbox{\gls{los} obstruction} & \ref{sec:configres:gain_attenuation} & Attenuation due to a person obstructing the \gls{los}. Lower values show that the configuration can benefit from alternative propagation paths. & The \gls{hmd} requires a large enough compound \gls{fov} to capture alternative \glspl{mpc} when the \gls{los} becomes obstructed. \\
    
    \rowcolor{pw_gray} RMS delay spread and \mbox{frequency-selective fading} & \ref{sec:configres:delay_spread} & A measure of the \gls{mpc} time delay differences and the severity of fading dips across the transmission bandwidth (ideally zero). & Configurations with different \glspl{fov} all experience a similar \gls{rms} delay spread, requiring similar channel equalization complexity. \\
    
    \rowcolor{wr_gray} Spatial multiplexing capability & \ref{sec:configres:spatial_multiplexing} & The capability of the array configuration to benefit from spatial multiplexing. The higher the capacity per additional stream, the better. & All except the single-array configuration (insufficient \gls{fov}) show significant capacity improvement due to spatial multiplexing.\\
    
    \rowcolor{pw_gray} Minimal service & \ref{sec:configres:minimal_service} & The minimal achievable channel capacity. Larger values indicate higher potential data rates, hence, better video quality. & A sufficient \gls{fov} (metric \ref{sec:configres:gain_tradeoff}) is essential for high single-stream capacity. Additional arrays can enhance multi-stream capacity.\\
    
    \bottomrule
\end{tabular}

\end{table*}

The position of the \gls{ap} plays an important role in optimizing signal coverage and enhancing network performance. For example, a ceiling\nobreakdash-mounted \gls{ap} is less susceptible to blockage, a corner\nobreakdash-mounted \gls{ap} may cover the entire floor area with a single array's \gls{fov}, and a wall\nobreakdash-mounted \gls{ap} can utilize additional walls as reflective surfaces. To evaluate the latter two hypotheses, we analyze and compare performance in \gls{los} and \gls{olos} conditions for either a corner\nobreakdash-mounted or a wall\nobreakdash-mounted \gls{ap}. To achieve this, we repeat part of the measurements (red\nobreakdash-white markers in \Cref{fig:measurement_environment_floorplan}) after moving the \gls{ap} out of the corner and placing it parallel to the nearby wall (blue\nobreakdash-white marker in \Cref{fig:measurement_environment_floorplan}). We then compare the mean channel gain, the time delay, and the azimuth spread between the two \gls{ap} deployments to establish the influence of the \glsposs{ap} position on channel performance. The \gls{rms} delay spread ($\sigma_\tau$) is calculated according to \Cref{eq:rms_delay_spread}, wherein the subtracted term represents the square of the mean excess delay ($\overline{\tau}^2$). We include the latter in the evaluation to provide an easier comparison between our measurements and the prior art. Azimuth spread is calculated according to \cite{cai_dynamic_2020} as
\begin{equation}
\label{eq:azimuth_spread}
 \sigma_{\phi} = \sqrt{ -2\log_{10}\left(\left|\frac
 {\sum\limits_{l=0}^{L-1}e^{j\phi_{\mathrm{R},l}}||\bm{\alpha}_l||_\mathrm{F}^2}
 {\sum\limits_{l=0}^{L-1}||\bm{\alpha}_l||_\mathrm{F}^2}
 \right|\right) },
\end{equation}
where $||\bm{\alpha}_l||_\mathrm{F}^2$ is the $l^{\text{th}}$ \glsposs{mpc} combined gain across all polarizations. A large $\sigma_{\phi}$ signifies that the \glspl{mpc} impinging on the receiver are highly spread out along the azimuth. Note that some prior art, such as \cite{hao_xu_spatial_2002}, applies a slightly different notation and the relationship to \Cref{eq:azimuth_spread} is $\sigma_\phi=\sqrt{-\log_{10}(1-\Lambda^2)}$, where $\Lambda$ is the alternative angular spread definition. We opted for \Cref{eq:azimuth_spread} due to its ability to assess differences in angular spread at large spread values, where $\Lambda$ would otherwise quickly converge towards one.

In this context, we also conduct an analysis of the \gls{los} \gls{ple} to characterise the measurement environment and verify the conformity of our data with the prior art \cite{hao_xu_spatial_2002, haneda_indoor_2016, cai_dynamic_2020, ju_millimeter_2021}. The \gls{ple} is derived by fitting a linear regression line to the gain of the first channel snapshot at each \gls{hmd} position, while minimizing the mean squared error. The \gls{ple} applies to the general path loss equation, noted below
\begin{equation}
\label{eq:path_loss}
    \mathrm{PL} (f,d) = \mathrm{PL}(f, d_0) \cdot \left( \frac{d}{d_0} \right)^{\!\!n} \! \cdot \mathcal{X}_\sigma,
\end{equation}
where $n$ is the \gls{ple}, $\mathrm{PL}(f, d_0)$ is the path loss at a reference distance $d_0$, and $\mathcal{X}_\sigma$ represents shadow fading ($\mathcal{X}_\sigma \geq 1$).

\section{Influence of antenna array configuration}
\label{sec:res_antenna_array_configuration}

This section applies the proposed performance metrics to channel measurements in order to approximate the performance of an \gls{hmd} featuring a variable number of antenna arrays. All metrics from \Cref{sec:performacne_metrics} are employed to show the trade\nobreakdash-offs in utilizing less antenna arrays than what would result in \SI{360}{}\nobreakdash-degree azimuth coverage. We focus on the full channel gain and minimum/mean service performance for both \glsentryfull{sm} and \glsentryfull{det}. We consider full $4\times4$ \gls{hmd} antenna arrays and the \gls{ap}$_0$ deployment for the purpose of the current study step. All gain and ratio values are from hereon presented on a logarithmic scale (in \SI{}{\decibel}). \Cref{tab:performance_metric_overview} provides a summary of the metrics and the insights they provide for \gls{mmw} \glspl{hmd}.

\subsection{Gain dependency on the azimuth FoV}
\label{sec:configres:gain_tradeoff}

\begin{figure}[h]
    \centering
    \plotFig{\plotFlag}{\resizebox{1\columnwidth}{!}{\input{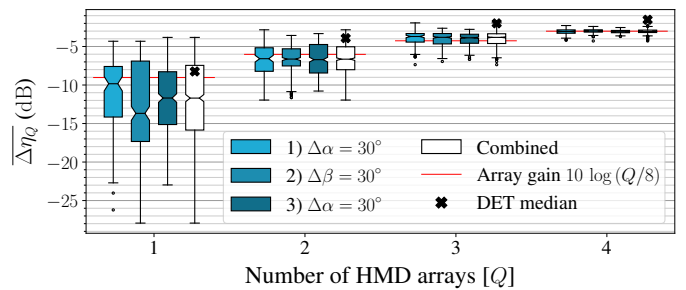}}}
    \caption{Gain ratio per mobility sequence part (blue) and combined (white). The middle line is the median ($\mathrm{P}_{{50}}$)), whiskers show $\mathrm{P}_{50}\pm1.5(\mathrm{P}_{75}-\mathrm{P}_{25})$, dots mark outliers, while notches show the \SI{95}{\percent} median confidence interval~\cite{mcgill_variations_1978}. The same box plot definition is applied throughout the paper.}
    \label{fig:gain_trade_off}
    \vspacebelowfig
\end{figure}

\Cref{fig:gain_trade_off} shows (from right to left) that deactivating half of the \gls{hmd} arrays, i.e., using four arrays, results in a gain loss proportional to the reduction in array gain (red line). A similar observation holds for a 3\nobreakdash-array configuration. The median gain for a 2\nobreakdash-array configuration decreases noticeably more than the array gain, indicating an insufficiently wide \gls{fov}. This becomes even more apparent for the single\nobreakdash-array configuration. The highly\nobreakdash-negative values for the single\nobreakdash-~and 2\nobreakdash-array configuration were recorded when the \gls{los} and the major \glspl{mpc} did not fall within the configuration's azimuth \gls{fov} due to \gls{hmd} rotation.

These observations contradict the \gls{det} gain results from our preceding work \cite{marinsek_impact_2023}, where the reduction in median \gls{det} gain ($\times$ mark in \Cref{fig:gain_trade_off}) is lesser than the reduction in array gain (red line). In the \gls{det} results, the benefits of a 4$^{\text{th}}$ array are also far less pronounced. We associate this with the fact that, in the presence of a \gls{los} component, \gls{det} beams primarily towards the \gls{los}, hence reducing the advantages of a multi\nobreakdash-array system. Lastly, from the blue boxes, we notice a single\nobreakdash-array \gls{hmd} performs better when standing upright than when tilted down, while the other configurations exhibit similar performance throughout the mobility sequence. This is due to the single array not receiving many prosperous \glspl{mpc} when angled down since the floor is covered with a carpet and often cluttered (table and chair legs). The latter also holds for \gls{det} gain, as demonstrated in our previous work \cite{marinsek_impact_2023}.

\vspace{-1mm}
\vspace{-2mm}

\subsection{Gain stability during rotation}
\label{sec:configres:gain_stability}

\begin{figure}[h]
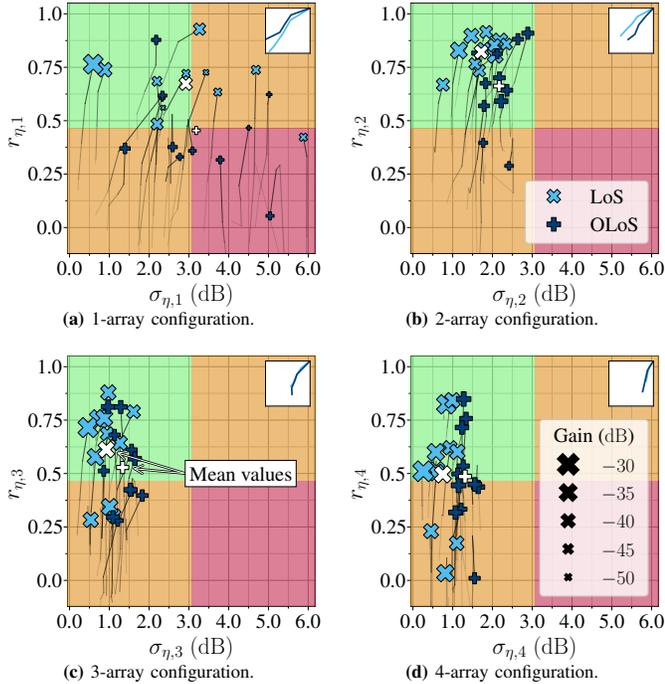

    \begin{subfigure}[b]{0.235\textwidth}
        \centering
        \plotFig{\plotFlag}{\resizebox{1\columnwidth}{!}{\input{figure-journal/gain_volatility_40.pgf}}}\vspace{-0.2cm}
        \caption{1-array configuration.}
        \label{fig:gain_volatility_1}
    \end{subfigure}
    \hfill
    \begin{subfigure}[b]{0.235\textwidth}
        \centering
        \plotFig{\plotFlag}{\resizebox{1\columnwidth}{!}{\input{figure-journal/gain_volatility_11.pgf}}}\vspace{-0.2cm}
        \caption{2-array configuration.}
        \label{fig:gain_volatility_2}
    \end{subfigure}
    \hfill
    \begin{subfigure}[b]{0.235\textwidth}
        \centering
        \plotFig{\plotFlag}{\resizebox{1\columnwidth}{!}{\input{figure-journal/gain_volatility_4a.pgf}}}\vspace{-0.2cm}
        \caption{3-array configuration.}
        \label{fig:gain_volatility_3}
    \end{subfigure}
    \hfill
    \begin{subfigure}[b]{0.235\textwidth}
        \centering
        \plotFig{\plotFlag}{\resizebox{1\columnwidth}{!}{\input{figure-journal/gain_volatility_aa.pgf}}}\vspace{-0.2cm}
        \caption{4-array configuration.}
        \label{fig:gain_volatility_4}
    \end{subfigure}\vspace{-0.2cm}
    \caption{Gain stability for the 11 \gls{hmd} positions and \gls{los}/\gls{olos}, based on the gain's standard deviation ($\sigma_{\scalebox{0.6}{$Q$}}$, horizontal) and autocorrelation ($r_{\scalebox{0.6}{$Q$}}$, vertical). The marker tails represent an increase in delay ($j$) from one to five samples, with the tail averages summarized in the top-right box of each plot.}
    \label{fig:gain_volatility}
\end{figure}

Gain stability during the course of a \SI{33}{\second} measurement is assessed in \Cref{fig:gain_volatility}. Standard deviation is used to evaluate the gain's spread, while autocorrelation shows its temporal consistency. The noticeable dispersion of points in \Cref{fig:gain_volatility_1} shows that a single\nobreakdash-array \gls{hmd} experiences both high gain spread and low gain consistency. The tails, starting from each point, show how the results change when increasing the autocorrelation delay from one sample ($j=1$) to five samples ($j=5$) and reducing the number of gain time series samples over which the standard deviation is computed by $2j$ (remove first and last $j$\nobreakdash-samples). For an easier overview and comparison against other array configurations, the mean tails are plotted at the top right of Figs.~\ref{fig:gain_volatility_1}\nobreakdash--\ref{fig:gain_volatility_4}. 
As can be seen in \Cref{fig:gain_volatility_2}, including a 2$^{\text{nd}}$ array increases gain consistency; however, the gain spread during the course of a measurement remains high. This is reduced by adding a 3$^{\text{rd}}$ array, as observed in \Cref{fig:gain_volatility_3}. A 4$^{\text{th}}$ array further decreases gain spread by a small amount, visible by comparing the mean markers (colored white) in Figs.~\ref{fig:gain_volatility_3} and~\ref{fig:gain_volatility_4}. Note that gain consistency increases with each additional \gls{hmd} array in spite of the lower autocorrelation ($r$ in \Cref{fig:gain_volatility}). The latter originates from a stable gain with slight variation, resembling a constant value with superimposed noise. Such outliers can be detected by either examining the time series data or by observing the vertical tails in \Cref{fig:gain_volatility_4}.

\vspace{-1mm}
 \vspace{-0.8cm}

\subsection{Attenuation due to LoS obstruction}
\label{sec:configres:gain_attenuation}

\begin{figure}[h]
    \centering
    \plotFig{\plotFlag}{\resizebox{1\columnwidth}{!}{\input{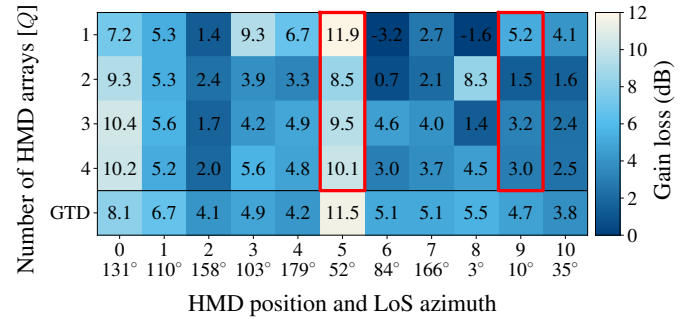}}}
    \caption{\gls{los}/\gls{olos} ratio for the 1$^{\text{st}}$ snapshot and the \gls{gtd} analytical reference. Red shows the positions where the \gls{hmd} was oriented towards the \gls{ap}.}
    \label{fig:obstruction_resillience_1st_snap}
    \vspacebelowfig
\end{figure}

The adverse effects of \gls{los} obstruction are presented in \Cref{fig:obstruction_resillience_1st_snap}. Here we focus primarily on positions \gls{hmd}$_5$ and \gls{hmd}$_9$, when the \gls{hmd} was oriented towards the \gls{ap} (highlighted in red). From the two, we notice that i) the effects of blockage on a single\nobreakdash-array \gls{hmd} are approx. \SI{2}{\decibel} more severe than for a 3\nobreakdash-array \gls{hmd} ii) a 3- and 4\nobreakdash-array configuration fare similarly in the presence of blockage, and iii) the adverse effects of \gls{los} obstruction noticeably decrease with distance, as the gain loss becomes as low as \SI{5}{\decibel} for position \gls{hmd}$_9$. The bottom row in \Cref{fig:obstruction_resillience_1st_snap} shows the reference gain loss, derived analytically using the \gls{gtd}. We consider the obstructing head can be approximated as a vertical cylinder with a \SI{15}{\centi\meter} diameter and that the \gls{olos} passes through the cylinder's central axis. The \gls{gtd} results confirm the diminishing attenuation with distance trend, and they align well with the results of a single\nobreakdash-array configuration when it is oriented towards the \gls{ap} (positions \gls{hmd}$_5$ and \gls{hmd}$_9$). However, we notice that the measured attenuation for a 4\nobreakdash-array configuration (full azimuth coverage) is several \SI{}{\decibel} smaller than the \gls{gtd} suggests. We explain this by the fact that the analytical model does not consider multipath propagation, which can reduce the attenuation caused by blockage. The negative values in \Cref{fig:obstruction_resillience_1st_snap} originate from comparisons where the \gls{hmd} had no clear \gls{los} due to poor orientation (misalignment) in neither the \gls{los} nor the \gls{olos} scenario. Recall that the orientations of the arrays in the 2\nobreakdash-array configuration are orthogonal to the array in a single\nobreakdash-array system. As a result, the effects of blockage on a 2\nobreakdash-array \gls{hmd} are more pronounced when the \gls{hmd} is oriented perpendicular to the \gls{ap}.

\vspace{-1mm}

\begin{figure}[h]
    \centering
    \plotFig{\plotFlag}{\resizebox{1\columnwidth}{!}{\input{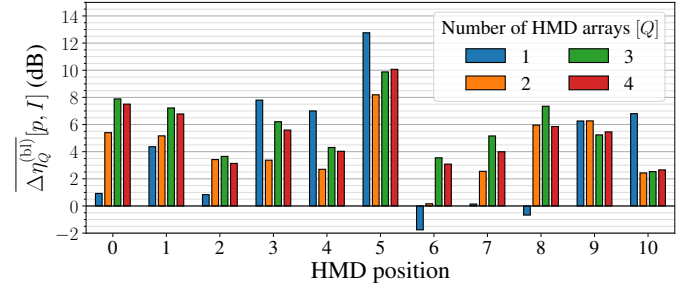}}}
    \caption{Ratio between mean \gls{los} and \gls{olos} gain. From left to right, the 4 columns at each \gls{hmd} position represent the 1-, 2-, 3-, and 4-array system.}
    \label{fig:obstruction_resillience_gc}
    \vspacebelowfig
\end{figure}

\Cref{fig:obstruction_resillience_gc} shows the ratio between mean \gls{los} and mean \gls{olos} gain. From it, we can confirm that the 3- and 4\nobreakdash-array configurations are similarly affected by blockage. The single\nobreakdash-array configuration features on average several decibels higher attenuation in cases where it was exposed to the \gls{los} and low/negative values when the array was in \gls{nlos} conditions during both the \gls{los} and \gls{olos} measurement. The 2\nobreakdash-array configuration shows consistent results with the 3- and 4\nobreakdash-array system, while it shows to be less affected by blockage. However, note that a 2\nobreakdash-array \gls{hmd} would experience lower mean gain as already shown in \Cref{fig:gain_trade_off}.

\vspace{-1mm}
\vspace{-2mm}

\subsection{RMS delay spread and frequency-selective fading}
\label{sec:configres:delay_spread}

\begin{figure}[h]
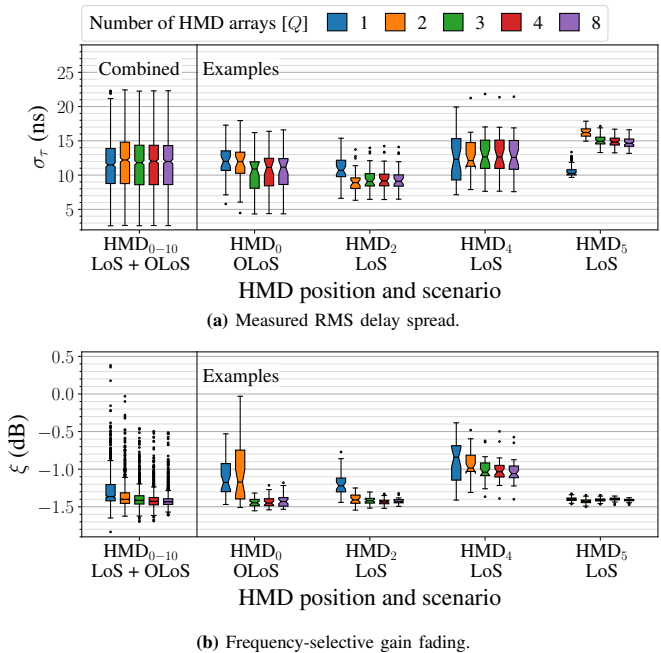

    \centering
    \begin{subfigure}[b]{0.99\columnwidth}
        \centering
        \plotFig{\plotFlag}{\resizebox{1\columnwidth}{!}{\input{figure-journal/rms_delay_spread_overview_selection.pgf}}}\vspace{-2mm}
        \caption{Measured \gls{rms} delay spread.}\vspace{-1mm}
        \label{fig:delay_spread}
    \end{subfigure}
    \hfill
    \begin{subfigure}[b]{0.99\columnwidth}
        \centering
        \begin{adjustbox}{clip, trim=0 0 0 5mm}
        \plotFig{\plotFlag}{\resizebox{1\columnwidth}{!}{\input{figure-journal/frequency_selective_fading_overview_selection.pgf}}}\vspace{-8mm}
        \end{adjustbox}
        \caption{Frequency-selective gain fading.}\vspace{-3mm}
        \label{fig:freq_sel_fading}
    \end{subfigure}
    \caption{Measured \gls{rms} delay spread and frequency-selective fading. Colors represent array configurations. The 8-array results are included for reference.}
    \label{fig:delay_spread_and_freq_sel_fading}
    \vspacebelowfig
\end{figure}

The reception of \glspl{mpc} and the flatness of the channel's spectrum are assessed in \Cref{fig:delay_spread_and_freq_sel_fading} through the \gls{rms} delay spread and frequency\nobreakdash-selective fading, respectively.
\Cref{fig:delay_spread} shows that all array configurations observe a similar \gls{rms} delay spread of \SIrange{2.5}{22.5}{\nano\second}, demonstrating that \glspl{mpc}, including higher order reflections, can contribute to channel gain. Moreover, the measured \gls{rms} delay spread translates to a coherence bandwidth of about \SIrange{45}{400}{\mega\hertz}. In addition to the combined result, \Cref{fig:delay_spread} illustrates the measurements for four \gls{hmd} positions to highlight the inconsistency of a single\nobreakdash-~and a 2\nobreakdash-array system. During \gls{olos} measurements at position \gls{hmd}$_0$, both configurations showed a large \gls{rms} delay spread since the \gls{olos} was no longer their dominant \gls{mpc}. The other configurations were more biased by the strong \gls{los} component in spite of the obstruction. Similarly, the single\nobreakdash-array configuration does not receive the \gls{los} at positions \gls{hmd}$_2$ and \gls{hmd}$_4$, while it is strongly influenced by higher order reflections. The single array receives a strong \gls{los} component and few other \glspl{mpc} at position \gls{hmd}$_5$, resulting in a low \gls{rms} delay spread. The 2\nobreakdash-array configuration is less influenced by the \gls{los} and more by the \glspl{mpc} at position \gls{hmd}$_5$ due to the orientation of the arrays, yielding a larger \gls{rms} delay spread.

\Cref{fig:freq_sel_fading} shows that, expectantly, combining the channels from a smaller number of arrays (fewer antennas) results in a less consistent gain across the measured bandwidth. Namely, the single\nobreakdash-~and 2\nobreakdash-array configurations feature noticeably worse performance. From the same examples as in \Cref{fig:delay_spread}, we see that higher \gls{rms} delay spread measurements are usually reflected in pronounced frequency selectivity. Therefore, the results in \Cref{fig:delay_spread_and_freq_sel_fading} indicate, that there are several \glspl{mpc} available in the channel in case of \gls{olos} conditions, however, due to the \glspl{mpc}, the channel can not be considered flat over the measured band and channel equalization needs to be applied.

\vspace{-2mm}
\vspace{-2mm}

\subsection{Spatial multiplexing capability}
\label{sec:configres:spatial_multiplexing}

\begin{figure}[h]
    \centering
    \plotFig{\plotFlag}{\resizebox{1\columnwidth}{!}{\input{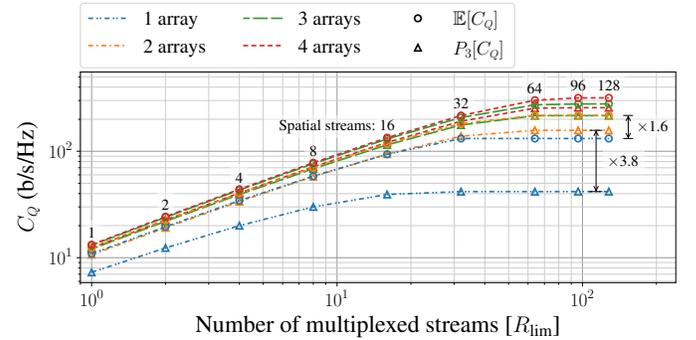}}}
    \caption{Stream-dependent service performance at an \gls{snr} of \SI{10}{\decibel}.}
    \label{fig:service_vs_num_of_streams}
    \vspacebelowfig
\end{figure}

\Cref{fig:service_vs_num_of_streams} shows the mean and minimal service, $\mathbb{E}\,[C_{\scalebox{0.6}{$Q$}}]$ and $P_3\,[C_{\scalebox{0.6}{$Q$}}]$, respectively, at an \gls{snr} of \SI{10}{\decibel} and for an increasing number of multiplexed streams. These are limited according to $R_{\scalebox{0.6}{$Q$}}\!\!=\!\!\min(M_{\scalebox{0.6}{$Q$}},N_{\scalebox{0.6}{$Q$}},R_{\text{lim}})$, where $R_{\text{lim}}$ is the maximal number of permitted streams. Starting from a single stream and \glsentryfull{det}, we see that minimal and mean service performance exhibit near\nobreakdash-linear growth on the logarithmic scale for \glsentryfull{sm} and up to 16~streams. The obtained capacity increases with the number of streams approximately according to $C \propto R_{\text{lim}}^{0.8}$. Hence, capacity per stream is highest for \gls{det}, and it begins to noticeably decline for more than 16 streams. Furthermore, \Cref{fig:service_vs_num_of_streams} shows that the ratio between the minimal service of a single\nobreakdash-array and that of a 2\nobreakdash-array configuration is larger than their mean service ratio (3.8 versus 1.6). A similar, yet less pronounced, observation can be made for the other configurations.

\vspace{-2mm}

\begin{figure}[h]
    \centering
    \plotFig{\plotFlag}{\resizebox{1\columnwidth}{!}{\input{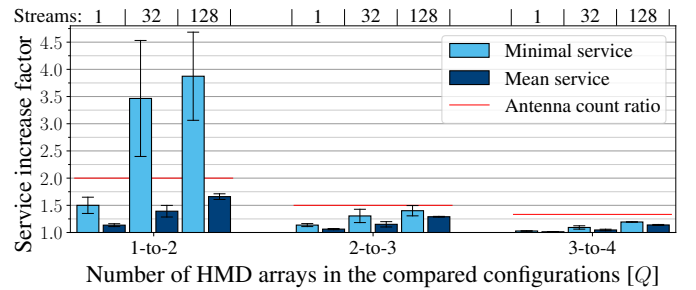}}}
    \caption{Service improvement due to an additional array, averaged across \glspl{snr} \SIrange{0}{20}{\decibel}. Vertical confidence intervals represent the standard deviation.}
    \label{fig:service_vs_num_of_panels}
    \vspacebelowfig
\end{figure}

\Cref{fig:service_vs_num_of_panels} plots the minimal and mean service ratios between 1~and~2, 2~and~3, and 3~and~4 arrays. The procedure is repeated for 1, 32, and 128 spatial streams. These represent \gls{det}, the maximal channel rank of the single\nobreakdash-array configuration (smallest channel matrix), and the maximal number of available eigenmodes for the 4\nobreakdash-array \gls{hmd} (largest matrix). Recall that the \gls{ap} features 128~antenna ports, enabling \gls{sm} on up to 128~streams. In \Cref{fig:service_vs_num_of_panels}, we can see that adding an antenna array to the \gls{hmd} yields a significantly more pronounced increase in the minimal service (light shade) than in the mean service (dark shade). Hence, \Cref{fig:service_vs_num_of_panels} highlights that evaluating only mean performance can lead to severely underestimating the benefits of employing additional arrays. The performance enhancements are more noticeable for a larger number of eigenmodes, since adding arrays broadens the \gls{fov}, which often yields additional \glspl{mpc}. 

\vspace{-1mm}
\vspace{-2mm}

\subsection{Minimal service}
\label{sec:configres:minimal_service}

\begin{figure}[h]
    \centering
    \begin{subfigure}[b]{0.99\columnwidth}
        \centering
        \plotFig{\plotFlag}{\resizebox{1\columnwidth}{!}{\begin{tikzpicture}[font=\fontsize{6}{6}\selectfont]

\node[anchor=south west,inner sep=0] (image) at (0,0) {
    \resizebox{1\columnwidth}{!}{\input{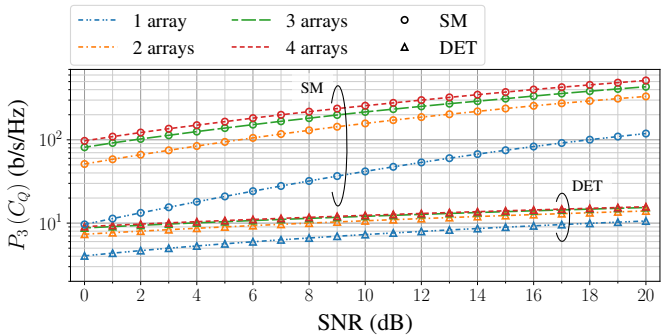}}
};

\begin{scope}[x={(image.south east)},y={(image.north west)}]
    \node at(0.465,0.745) [fill=white, opacity=.6, text opacity=1, anchor=center, align=center, text centered] {SM};
    \draw (0.495,0.436) arc [
        start angle=230,
        end angle=495,
        x radius=1.2mm,
        y radius=8.0mm
    ];
    \node at(0.88,0.455) [fill=white, opacity=.6, text opacity=1, anchor=center, align=center, text centered] {DET};
    \draw (0.833, 0.315) arc [
        start angle=220,
        end angle=500,
        x radius=1mm,
        y radius=3.2mm
    ];
\end{scope}

\end{tikzpicture}%
}}\vspace{-1mm}
        \caption{Channel gain first averaged across the transmission bandwidth (\SI{768}{\mega\hertz}).}
        \label{fig:minimal_service_vs_snr_sc}
    \end{subfigure}
    \hfill
    \begin{subfigure}[b]{0.99\columnwidth}
        \centering
        \begin{adjustbox}{clip, trim=0 0 0 8.6mm}
        \plotFig{\plotFlag}{\resizebox{1\columnwidth}{!}{
\begin{tikzpicture}[font=\fontsize{6}{6}\selectfont]

\node[anchor=south west,inner sep=0] (image) at (0,0) {
    \resizebox{1\columnwidth}{!}{\input{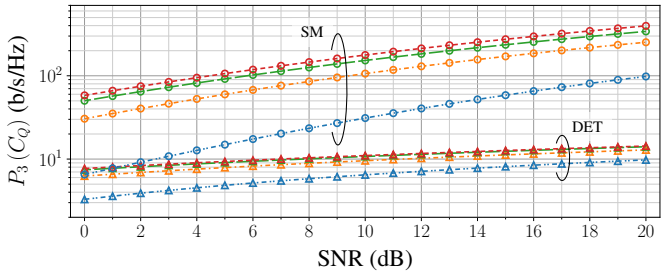}}
};

\begin{scope}[x={(image.south east)},y={(image.north west)}]
    \node at(0.465,0.72) [fill=white, opacity=.6, text opacity=1, anchor=center, align=center, text centered] {SM};
    \draw (0.495,0.424) arc [
        start angle=230,
        end angle=495,
        x radius=1.2mm,
        y radius=7.4mm
    ];
    \node at(0.88,0.44) [fill=white, opacity=.6, text opacity=1, anchor=center, align=center, text centered] {DET};
    \draw (0.833, 0.304) arc [
        start angle=220,
        end angle=500,
        x radius=1mm,
        y radius=3mm
    ];
\end{scope}

\end{tikzpicture}%
}}
        \end{adjustbox}\vspace{-1mm}
        \caption{Capacity calculated per sounding tone (\SI{375}{\kilo\hertz} tone spacing).}
        \label{fig:minimal_service_vs_snr_ofdm}
    \end{subfigure}
    \vspace{-0.7 cm}
    \caption{Minimal service for \gls{sm} and \gls{det}.}
    \label{fig:minimal_service_vs_snr}
\end{figure}

\Cref{fig:minimal_service_vs_snr_sc} shows the obtained minimal service ($P_3\,[C_{\scalebox{0.6}{$Q$}}]$) for a 1--4\nobreakdash-array configuration, employing either \gls{sm} or \gls{det} and averaging the eigenmode gains across the \SI{768}{\mega\hertz} bandwidth. For \gls{det}, we set $R_{\scalebox{0.6}{$Q$}}=1$ in \Cref{eq:minimal_service} in order to limit the system to a single eigenmode/stream. Both blue curves confirm the previous observations, that a single\nobreakdash-array \gls{hmd} is severely affected by rotation. The 2\nobreakdash-array configuration improves capacity, however, it still noticeably under\nobreakdash-performs when compared to the 3- and 4\nobreakdash-array configuration. The latter two offer comparable capacity for \gls{det}, while a somewhat larger performance factor separates them for \gls{sm}. This indicates that a 3\nobreakdash-array \gls{hmd} can capture most of the available signal components, while a fully\nobreakdash-digital 4\nobreakdash-array \gls{mimo} system further enhances performance using the additional multiplexed streams. The dual\nobreakdash-polarized 1, 2, 3, and 4~arrays can utilize up to 32, 64, 96, and 128~multiplexed streams, respectively. \Cref{fig:minimal_service_vs_snr_ofdm} shows that the minimal service is lower when calculated per sounding tone, since certain parts of the transmission bandwidth can suffer from fading. Regardless, we observe the same trend as in \Cref{fig:minimal_service_vs_snr_sc}~\nobreakdash--~a large gap between a single\nobreakdash-~and a 2\nobreakdash-array configuration, which narrows as a 3$^{\text{rd}}$ and a 4$^{\text{th}}$ array is added.

\vspace{-2mm}
\vspace{-0.5 cm}

Recall that capacity is calculated for normalized channels and that the evaluated channel matrices consist of \SIrange{4096}{16384}{} individual channels. Hence, the relatively high capacity figures. We consider that all configurations have an equal transmit power budget in \Cref{fig:minimal_service_vs_snr}. Solely from this perspective, the 4\nobreakdash-array \gls{hmd} that utilizes \gls{sm} is the most energy efficient (lowest energy per transmitted bit). However, powering the additional hardware for each array and utilizing complex baseband processing is likely to reduce energy efficiency in practice. An analysis of the energy efficiency of individual components would be required to reach further conclusions.

\section{Additional degrees of freedom}
\label{sec:res_additional_considerations}

In this section, we examine the impact of: A.~equipping the \gls{hmd} with a rear headband and mounting an array to it, B.~varying the number of active antennas per \gls{hmd} antenna array, and C.~placing the \gls{ap} on a wall instead of in the room's corner. Namely, we evaluate the channel in terms of the six performance metrics and the procedures introduced in Sections \ref{sec:aproc:read_headband}\nobreakdash--\ref{sec:aproc:ap_position}. Only the most insightful results are presented.

\subsection{Utilizing a rear headband}
\label{sec:addconsres:res_rear_headband}

\begin{figure}[h]
    \centering
    \plotFig{\plotFlag}{\resizebox{1\columnwidth}{!}{
\begin{tikzpicture}[font=\fontsize{6}{6}\selectfont]

\node[anchor=south west,inner sep=0] (image) at (0,0) {
    \resizebox{1\columnwidth}{!}{\input{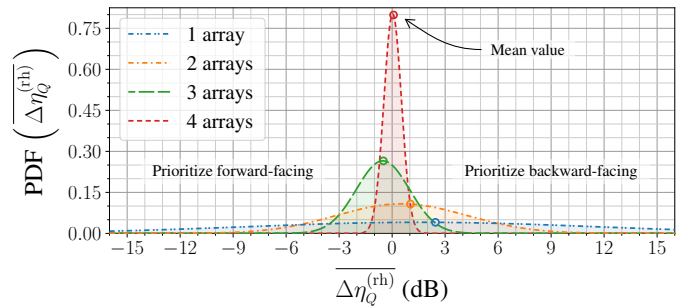}}
};

\begin{scope}[x={(image.south east)},y={(image.north west)}]

    \fill [white, opacity=.6] (0.70, 0.82) rectangle ++(0.14, 0.06);
    \node (h) at(0.77, 0.85) [anchor=center, align=center, text centered] {Mean value};
    \draw[black, -{angle 60}] (h.west) to[out=180, in=-34] (0.586, 0.952);

\end{scope}

\end{tikzpicture}%
}}
    \caption{\Glsentryfull{pdf} of the gain ratio between backward- and forward-facing $Q$-array configurations.}
    \label{fig:rear_headband_pdf}
    \vspacebelowfig
\end{figure}

\Cref{fig:rear_headband_pdf} shows the gain ratio between a forward\nobreakdash-facing and a backward\nobreakdash-facing $Q$\nobreakdash-array configuration. The ratios are represented by fitted Gaussian distributions for easier viewing. We notice that a single\nobreakdash-array setup facing backward results in a higher average gain (positive mean value in \Cref{fig:rear_headband_pdf}). This means that orienting the array towards the ceiling, rather than the floor, leads to increased gain since the \gls{hmd} is primarily pitched downward (forward) in our mobility sequence. We attribute the lower gain when pointing towards the floor to the presence of a carpeted and cluttered floor (chair and table legs). Similarly, the 2\nobreakdash-array setup favors the backward\nobreakdash-facing configuration (arrays at the back and front instead of above the user's ears), which can be partially attributed to the applied mobility pattern. This pattern benefits the backward\nobreakdash-facing configuration when the \gls{hmd} pitches towards or away from the access point (the pitching motion accounts for approx. 15 out of the total \SI{33}{\second}). On the other hand, a 3\nobreakdash-array \gls{hmd} performs better in the forward\nobreakdash-facing configuration (negative mean value in \Cref{fig:rear_headband_pdf}), where two arrays are diagonally positioned at the back and only one array is placed at the front. Hence, the configuration emphasizes \glspl{mpc} directed towards the ceiling. A 4\nobreakdash-array system does not show a preference for either the forward- or backward\nobreakdash-facing configuration. 

\vspace{-1mm}

We expect that the observed biases for the \SIrange{1}{3}{}\nobreakdash-array configurations could become less apparent, or even of opposite value, in environments where the floor provides more reflections. Note that the compactness and offset of the \glspl{pdf} demonstrate the capabilities of individual array configurations in capturing the available \glspl{mpc}.

\subsection{Activating a limited number of antennas in HMD arrays}
\label{sec:addconsres:deactivating_antennas}

\begin{figure}[h]
    \centering
    \plotFig{\plotFlag}{\resizebox{1\columnwidth}{!}{\input{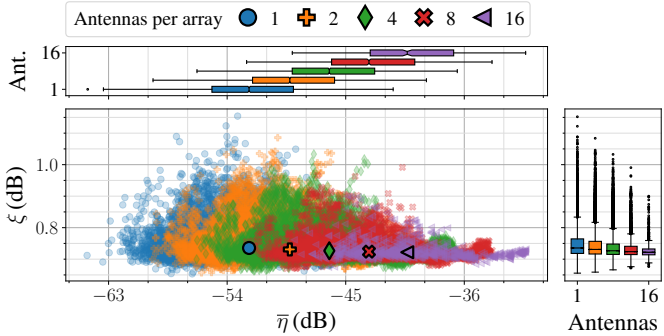}}}\vspace{-1mm}
    \caption{Gain (x-axis) and frequency-selective fading (y-axis) for a \mbox{3-array \gls{hmd}} configuration with 1--16 active antennas per array.}
    \label{fig:gain_vs_frequency_persistence}
\end{figure}

We have chosen to evaluate the impact of \gls{hmd} antenna deactivation on the 3\nobreakdash-array configuration since it offers a good balance between performance and the number of arrays, as observed in \Cref{sec:res_antenna_array_configuration}. \Cref{fig:gain_vs_frequency_persistence} shows that activating more antennas results in a gain increase corresponding to the larger array gain (x\nobreakdash-axis), while the gain's distribution becomes more compact as more antennas are activated. The additional antennas increase spatial diversity, lowering the likelihood that a major portion of the channels suffers from fading simultaneously, both across the measured bandwidth (frequency) and across the multiple channel snapshots (time). Correspondingly, activating a larger number of antennas contributes to lower and more compactly distributed frequency\nobreakdash-selective fading (y\nobreakdash-axis).

\vspace{-2mm}
\vspace{-4mm}

\begin{figure}[h]
    \plotFig{\plotFlag}{\resizebox{1\columnwidth}{!}{\begin{tikzpicture}[font=\fontsize{6}{6}\selectfont]

\node[anchor=south west,inner sep=0] (image) at (0,0) {
    \resizebox{1\columnwidth}{!}{\input{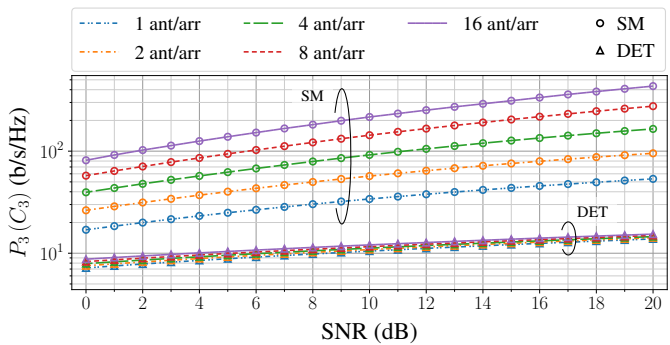}}
};

\begin{scope}[x={(image.south east)},y={(image.north west)}]
    \node at(0.46,0.74) [fill=white, opacity=.6, text opacity=1, anchor=center, align=center, text centered] {SM};
    \draw (0.495,0.41) arc [
        start angle=230,
        end angle=500,
        x radius=1.2mm,
        y radius=9.0mm
    ];
    \node at(0.878,0.40) [fill=white, opacity=.6, text opacity=1, anchor=center, align=center, text centered] {DET};
    \draw (0.833, 0.290) arc [
        start angle=220,
        end angle=510,
        x radius=1mm,
        y radius=2.2mm
    ];
\end{scope}

\end{tikzpicture}%
}}\vspace{-1mm}
    \caption{Minimal service for 3-array \gls{hmd}, 1--16 active antennas per array.}\vspace{-1mm}
    \label{fig:capacity_antenna_deactivation}
\end{figure}

\Cref{fig:capacity_antenna_deactivation} shows that minimal service (capacity's 3$^{\text{rd}}$ percentile) increases according to the array gain as more antennas are activated. Contrary to the prior evaluation of the number of active antenna arrays in \Cref{sec:configres:minimal_service}, all of the \gls{sm} curves in \Cref{fig:capacity_antenna_deactivation} bear a similar shape. This is because a three\nobreakdash-array configuration can detect the same \glspl{mpc} using its (beamformed) \gls{fov}, regardless of the number of active antennas. For example, if three arrays, each with a single active antenna, receive the \gls{los} component, then so will the three arrays with 16 active antennas per array. A larger number of active antennas does, however, increase the beamforming capabilities of the system. The improved spatial filtering enhances \gls{sm} capabilities, resulting in a consistent capacity gap even as \gls{snr} increases. An exception are the \gls{det} curves, which start to converge for higher \glspl{snr} due to the logarithmic term in calculating capacity and the usage of a single stream. 

\vspace{-1mm}
\vspace{-4mm}

\begin{figure}[h]
    \centering
    \plotFig{\plotFlag}{\resizebox{1\columnwidth}{!}{
\begin{tikzpicture}[font=\fontsize{6}{6}\selectfont]

\node[anchor=south west,inner sep=0] (image) at (0,0) {
    \resizebox{1\columnwidth}{!}{\input{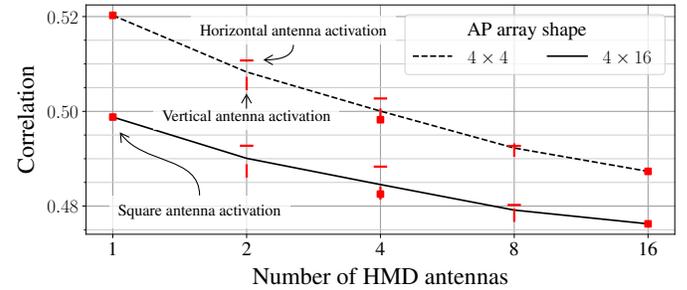}}
};

\begin{scope}[x={(image.south east)},y={(image.north west)}]

    \node (h) at(0.42,0.90) [fill=white, opacity=.8, text opacity=1, anchor=center, align=center, text centered] {Horizontal antenna activation};
    \draw[black, -{angle 60}] (h.south) to[out=-90, in=0] ([xshift=-4mm,yshift=-2mm]h.south);
    
    \node (v) at(0.35,0.60) [fill=white, opacity=.8, text opacity=1, anchor=center, align=center, text centered] {Vertical antenna activation};
    \draw[black, -{angle 60}] ([yshift=-1.1mm]v.north) to[out=90, in=-90] ([xshift=0mm,yshift=+0.9mm]v.north);
    
    \node (s) at(0.28,0.26) [fill=white, opacity=.8, text opacity=1, anchor=center, align=center, text centered] {Square antenna activation};
    \draw[black, -{angle 60}] (s.north) to[out=90, in=-55] ([xshift=-10.6mm,yshift=8.4mm]s.north);
    
\end{scope}

\end{tikzpicture}%
}}\vspace{-1mm}
    \caption{Mean channel gain correlation. Black curves represent the mean correlation across all antenna activation patterns for a given number of active antennas. Red vertical lines, horizontal lines, and squares show the mean correlation for vertical, horizontal, and square activation patterns, respectively.}
    \label{fig:antenna_deactivation_channel_correlation}
\end{figure}

\Cref{fig:antenna_deactivation_channel_correlation} shows the channel gain correlation when the \gls{ap} employs either a $4\times4$ (dashed) or a $4\times16$ (solid) array. The former is included to evaluate how correlation might change for smaller \gls{ap} arrays. For example, due to size and complexity constraints in practical deployments. \Cref{fig:antenna_deactivation_channel_correlation} confirms that channel gain correlation decreases with the number of active antennas. Hence, explaining the larger gain spread for configurations with less active antennas in \Cref{fig:gain_vs_frequency_persistence}. Furthermore, gain correlation is slightly larger across antennas in the horizontal than in the vertical direction. This happens as each array receives a portion of the available \glspl{mpc} due to a limited \gls{fov}, while, when a high\nobreakdash-gain \gls{mpc} gets reflected from the floor, a table, or the ceiling, a fading dip may occur due to destructive interference. Having a vertical array structure decreases the proportion of antennas that experience this fading dip. The somewhat smaller values of the curve corresponding to a $4\times16$ \gls{ap} array shows that reducing the likelihood of obtaining low gain due to simultaneous channel fading can be tackled by increasing either the number of \gls{hmd} or \gls{ap} antennas.

\vspace{-1mm}

\vspace{-0.8cm}

\subsection{Impact of AP position and path loss assessment}
\label{sec:addconsres:impact_of_ap_position}

\begin{figure}[h]
    \centering
    \plotFig{\plotFlag}{\resizebox{1\columnwidth}{!}{
\begin{tikzpicture}[font=\fontsize{6}{6}\selectfont]

\definecolor{gray1}{RGB}{220,220,220}
\definecolor{blue1}{RGB}{0,64,122}

\node[anchor=south west,inner sep=0] (image) at (0,0) {
    \resizebox{1\columnwidth}{!}{\input{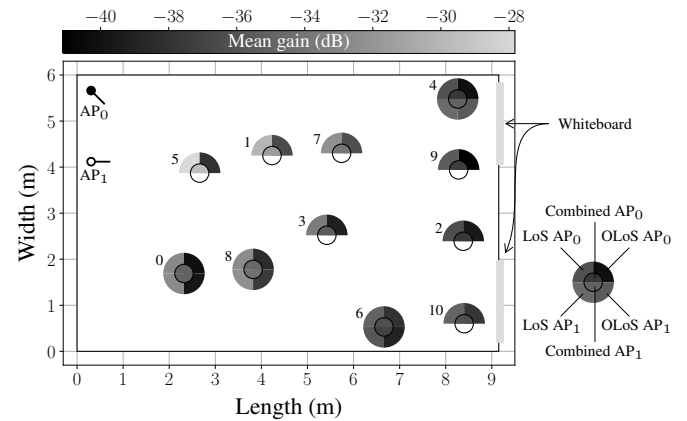}}
};

\begin{scope}[x={(image.south east)},y={(image.north west)}]

    
    \draw [gray1, line width = 3.00] (0.74, 0.62) -- ++ (0.0,0.2);
    \draw [gray1, line width = 3.00] (0.74, 0.19) -- ++ (0.0,0.2);
    \node (annotation) at (0.884,0.72) [anchor=center, align=center, text centered] {Whiteboard};
    \draw [black, line width = 0.3, -{angle 60}] (annotation.west) [out=180, in=0] to ([xshift=-5.9mm]annotation.west);
    \draw [black, line width = 0.3, -{angle 60}] (annotation.west) [out=180, in=65] to ([xshift=-5.9mm,yshift=-17.5mm]annotation.west);

    
    \coordinate (A) at (0.884,0.337);
    
    \draw [black, line width = 0.3] ([yshift=+0.6mm]A) -- ++(90:7.5mm);
    \draw [black, line width = 0.3] ([yshift=-0.6mm]A) -- ++(270:7.5mm);
    \draw [black, line width = 0.3] ([xshift=+1.6mm, yshift=+1.6mm]A) -- ++(45:4.3mm);
    \draw [black, line width = 0.3] ([xshift=-1.6mm, yshift=+1.6mm]A) -- ++(135:4.3mm);
    \draw [black, line width = 0.3] ([xshift=-1.6mm, yshift=-1.6mm]A) -- ++(225:4.3mm);
    \draw [black, line width = 0.3] ([xshift=+1.6mm, yshift=-1.6mm]A) -- ++(315:4.3mm);
    
    \node at ([yshift=+9.5mm]A) [anchor=center, align=center, text centered] {Combined AP$_0$};
    \node at ([yshift=-9.5mm]A) [anchor=center, align=center, text centered] {Combined AP$_1$};
    \node at ([xshift=+5.7mm, yshift=+6mm]A) [anchor=center, align=center, text centered] {OLoS AP$_0$};
    \node at ([xshift=-5.7mm, yshift=+6mm]A) [anchor=center, align=center, text centered] {LoS AP$_0$};
    \node at ([xshift=-5.7mm, yshift=-6mm]A) [anchor=center, align=center, text centered] {LoS AP$_1$};
    \node at ([xshift=+5.7mm, yshift=-6mm]A) [anchor=center, align=center, text centered] {OLoS AP$_1$};

    
    \fill[white] (0.1,0.55) rectangle ++(0.06,0.28);

    \coordinate (B) at (0.12, 0.80);
    \draw[black, line width = 0.7] (B) -- ++ (-45:2.6mm);
    \fill[fill=black] (B) circle (0.65mm);
    \node[black, anchor=center, align=center, text centered] at ([xshift=0.5mm, yshift=-3mm]B) {AP$_0$};

    \coordinate (C) at (0.12, 0.628);
    \draw[black, line width = 0.7] (C) -- ++ (0:2.6mm);
    \fill[fill=black] (C) circle (0.65mm);
    \fill[fill=white] (C) circle (0.4mm);
    \node[black, anchor=center, align=center, text centered] at ([xshift=0.5mm, yshift=-2mm]C) {AP$_1$};
    
\end{scope}

\end{tikzpicture}%
}}
    \caption{Mean spatial channel gain for both \gls{ap} deployments. The upper half-circles are reserved for \gls{ap}$_0$ and the lower for \gls{ap}$_1$. The left and right side show \gls{los} and \gls{olos} gain, respectively, with the combined gain in the center.}
    \label{fig:spatial_gain}
    \vspacebelowfig
\end{figure}

\Cref{fig:spatial_gain} illustrates the mean channel gain per \gls{hmd} position and scenario (\gls{los}/\gls{olos}/combined), and \gls{ap} deployment. The evaluation utilizes all 8~\gls{hmd} antenna arrays to provide a holistic view of the available channel gain. We notice that positions \gls{hmd}$_0$, \gls{hmd}$_6$, and \gls{hmd}$_8$ show similar gain for both \gls{ap}$_0$ and \gls{ap}$_1$, while \gls{hmd}$_4$ exhibits higher gain for \gls{ap}$_1$. Gain performance is, on average across the 4 common 
\gls{hmd} positions, in favor of \gls{ap}$_1$ by -0.3, 1.5, and \SI{0.6}{\decibel} for \gls{los}, \gls{olos}, and combined gain, correspondingly. 

\vspace{-0.2mm}

\Cref{tab:addconsres:time_and_angle_characterisation} shows that the higher gain during \gls{los} obstruction at position \gls{hmd}$_4$ for \gls{ap}$_1$ occurs together with a larger delay ($\sigma_\tau$) and angular spread ($\sigma_\phi$). This can be attributed to additional reflections from the longitudinal wall adjacent to \gls{hmd}$_4$ and the whiteboard behind it. The wall\nobreakdash-mounted \gls{ap} shows a slightly higher delay and angular spread on average (last row in \Cref{tab:addconsres:time_and_angle_characterisation}), however, further measurements would be required to conclude on optimal \gls{ap} deployment strategies. The time delay measurements in \Cref{tab:addconsres:time_and_angle_characterisation} exhibit a high degree of similarity to the reported values by prior art in an office environment and a small meeting room \cite{hao_xu_spatial_2002, wu_60-ghz_2017}. Namely, the similar room geometry is reflected in similar multipath propagation. Larger indoor environments, especially in \gls{olos} conditions, feature a larger delay and wider angular spread. For instance, $\sigma_\tau \approx \;$\SI{30}{\nano\second} and $\sigma_\phi \approx 1.5$ in a large lobby \cite{cai_dynamic_2020}. Outdoor scenarios in general feature larger dimensions with less scatterers, resulting in $\sigma_\tau > \;$\SI{30}{\nano\second} and $\sigma_\phi < 0.5$ \cite{hao_xu_spatial_2002}.

\vspace{-3mm}
\begin{table}[h]
    \centering
    \renewcommand{\arraystretch}{1.5}
    \caption{The channel's mean excess delay ($\overline{\tau}$), \gls{rms} delay spread ($\sigma_\tau$), and azimuth spread ($\sigma_\phi$). The final two rows contain the mean results. All~of~the presented results are derived using the full 8-array configuration.}
    \label{tab:addconsres:time_and_angle_characterisation}

\begin{tabularx}{\linewidth}{
>{\centering\arraybackslash}p{0.07\linewidth}>{\centering\arraybackslash}p{0.042\linewidth}||
>{\hfill}p{0.09\linewidth}|>{\hfill}p{0.09\linewidth}|>{\hfill}p{0.06\linewidth}||
>{\hfill}p{0.09\linewidth}|>{\hfill}p{0.09\linewidth}|>{\hfill}p{0.06\linewidth}
}
    \toprule
    
    \multicolumn{2}{c||}{Position} & \multicolumn{3}{c||}{Measurement -- \gls{los}} & \multicolumn{3}{c}{Measurement -- \gls{olos}} \\
    HMD & AP
    & $\overline{\tau}\,$(ns) & $\sigma_{\tau}\,$(ns) & $\sigma_{\phi}$ 
    & $\overline{\tau}\,$(ns) & $\sigma_{\tau}\,$(ns) & $\sigma_{\phi}$ \\
    
    \midrule
    
    \multirow{2}{*}{0} 
    & 0 & \cellcolor{wr_gray} 17.0 & \cellcolor{wr_gray}  7.7 & \cellcolor{wr_gray} 0.42 
    & \cellcolor{wr_gray} 27.7 & \cellcolor{wr_gray} 14.4 & \cellcolor{wr_gray} 0.71 \\
    & 1 & \cellcolor{pw_gray} 14.1 & \cellcolor{pw_gray}  8.1 & \cellcolor{pw_gray} 0.31 
    & \cellcolor{pw_gray} 22.6 & \cellcolor{pw_gray}  9.6 & \cellcolor{pw_gray} 0.50 \\
    
    \hline
    
    \multirow{2}{*}{4} 
    & 0 & \cellcolor{wr_gray} 30.9 & \cellcolor{wr_gray}  9.2 & \cellcolor{wr_gray} 1.06 
    & \cellcolor{wr_gray} 33.9 & \cellcolor{wr_gray}  9.5 & \cellcolor{wr_gray} 1.03 \\
    & 1 & \cellcolor{pw_gray} 33.3 & \cellcolor{pw_gray} 11.6 & \cellcolor{pw_gray} 1.21
    & \cellcolor{pw_gray} 36.1 & \cellcolor{pw_gray} 13.7 & \cellcolor{pw_gray} 1.33 \\
    
    \hline
    
    \multirow{2}{*}{6} 
    & 0 & \cellcolor{wr_gray} 29.6 & \cellcolor{wr_gray}  5.1 & \cellcolor{wr_gray} 0.26 
    & \cellcolor{wr_gray} 30.3 & \cellcolor{wr_gray}  7.2 & \cellcolor{wr_gray} 0.40 \\
    & 1 & \cellcolor{pw_gray} 29.0 & \cellcolor{pw_gray} 9.8 & \cellcolor{pw_gray} 0.55
    & \cellcolor{pw_gray} 33.3 & \cellcolor{pw_gray} 10.9 & \cellcolor{pw_gray} 0.78 \\
    
    \hline
    
    \multirow{2}{*}{8} 
    & 0 & \cellcolor{wr_gray} 21.4 & \cellcolor{wr_gray}  9.9 & \cellcolor{wr_gray} 0.46 
    & \cellcolor{wr_gray} 24.2 & \cellcolor{wr_gray} 10.7 & \cellcolor{wr_gray} 0.65 \\
    & 1 & \cellcolor{pw_gray} 18.1 & \cellcolor{pw_gray} 7.8 & \cellcolor{pw_gray} 0.36
    & \cellcolor{pw_gray} 23.6 & \cellcolor{pw_gray} 9.5 & \cellcolor{pw_gray} 0.44 \\
    
    \midrule
    
    \multirow{2}{*}{$\mathbb{E}$} 
    & 0 & \cellcolor{wr_gray} 24.7 & \cellcolor{wr_gray}  8.0 & \cellcolor{wr_gray} 0.55
    & \cellcolor{wr_gray} 29.0 & \cellcolor{wr_gray} 10.4 & \cellcolor{wr_gray} 0.70 \\
    & 1 & \cellcolor{pw_gray} 23.6 & \cellcolor{pw_gray} 9.3 & \cellcolor{pw_gray} 0.61
    & \cellcolor{pw_gray} 28.6 & \cellcolor{pw_gray} 10.9 & \cellcolor{pw_gray} 0.76 \\
    
    \bottomrule
\end{tabularx}

\end{table}

\begin{figure}[h]
    \centering
    \plotFig{\plotFlag}{\resizebox{1\columnwidth}{!}{\input{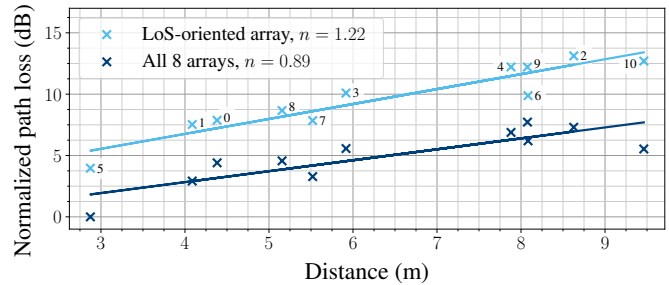}}}
    \caption{Normalized path loss. The two lines are derived using linear regression with squared error minimization. Light blue and dark blue values are normalized with the same weight in order to preserve the difference between them. Numbers represent \gls{hmd} positions.}
    \label{fig:spatial_gain_ple}
    \vspacebelowfig
\end{figure}

\Cref{fig:spatial_gain_ple} shows the path loss for the entire 8\nobreakdash-array configuration (dark shade) and the \gls{los}\nobreakdash-oriented array (light shade) at each \gls{hmd} position. Recall that, at each of the 11~positions, the \gls{hmd} had a random orientation. At each position, we select the array that is closest to boresigth and plot its path loss in light blue. This array receives the largest amount of power from the \gls{los} component among the eight \gls{hmd} arrays, while \gls{mpc} reception is limited by its constrained \gls{fov}. We see that the \gls{los}\nobreakdash-oriented array has a relatively low \gls{ple} ($n=1.22$) compared to prior art, which reports a \gls{ple} of $n=$\SIrange{1.7}{1.9}{} \cite{hao_xu_spatial_2002, haneda_indoor_2016, cai_dynamic_2020, ju_millimeter_2021}. We associate this with favorable \gls{mpc} propagation, including higher order reflections. The \glspl{mpc} are to a larger extent contained within the room due to the good reflective properties of the whiteboards, metal\nobreakdash-coated windows, and concrete walls \cite{zhao_28_2013}. The \gls{ple} becomes even smaller for the 8\nobreakdash-array configuration, which can receive \glspl{mpc} that impinge the \gls{hmd} from all azimuth directions. The smaller \gls{ple} in the latter case supports the findings on diminishing obstruction severity as distance increases, reported in \Cref{sec:configres:gain_attenuation}. Similar observations, although with somewhat larger \glspl{ple}, were made in \cite{ju_millimeter_2021}, where the authors reported $n=1.7$ and $n=1.2$ for the so\nobreakdash-called directional and omnidirectional \gls{ple}, respectively. Note that the vertical difference between the two fitted lines in \Cref{fig:spatial_gain_ple} represents the gain ratio between a single\nobreakdash-~and an 8\nobreakdash-array configuration in \gls{los} conditions.

\vspace{-1mm}
\vspace{0mm}

\section{Conclusion}
\label{sec:conclusion}
In this work, we studied different \gls{mmw} antenna array configurations for \glspl{hmd}. We introduced six channel performance metrics, analyzed them using analytical models and discussed their application\nobreakdash-level significance. Furthermore, the metrics were employed in an empirical evaluation, conducted on \SI{28}{\giga\hertz} \gls{mimo} indoor channel sounding data.

The results show that \gls{mmw} technology is suitable for high\nobreakdash-performance \gls{xr} applications when the \gls{hmd} receives the channel's major \glspl{mpc}. However, the rotation exhibited by \gls{hmd} users means that a single\nobreakdash-array system will observe unstable channel gain due to its limited azimuth \gls{fov}. Distributing additional arrays along the \glsposs{hmd} azimuth broadens its \gls{fov}, allowing for better \gls{mpc} reception and significantly stabilizing channel gain. Note, however, that the increase in gain is often smaller than suggested by theoretical array gain, since not all of the arrays are equally illuminated by impinging \glspl{mpc}. Attenuation upon obstruction by a person and its dependency on distance can be roughly predicted by the \glsposs{gtd} analytical model. Moreover, we noticed that the additional \glspl{mpc}, received by a multi\nobreakdash-array \gls{hmd}, can lessen the incurred attenuation by several~\SI{}{\decibel}. Finally, the reception of additional \glspl{mpc} can increase orthogonality among the channel's eigenmodes, enhancing the intrinsic capability of the channel to multiplex spatial streams. The results lead us to the conclusion that it is required to either equip the \gls{hmd} with multiple arrays or, alternatively, feature macroscopic diversity by employing distributed \gls{ap} infrastructure. The two can also be considered jointly for high\nobreakdash-reliability communication.

With the increased popularity of \gls{xr} and the pursuit towards wireless \glspl{hmd}, we can foresee a greater need in emulating physical layer behavior for \glspl{hmd}. Hence, an prosperous research direction for future work could be to integrate \gls{hmd} mobility data into channel models. Furthermore, the proposed performance metrics and methodology in this work can be applied to measurement campaigns in other environments, such as city streets and entertainment venues, to further evaluate the behavior of \gls{mmw} \glspl{hmd} in everyday settings.

\section*{Acknowledgement}

\vspace{1 mm}

\noindent
\footnotesize{Thanks to Meifang Zhu, Gilles Callebaut, and the Lund University volunteers.}
\vspace{-2 mm}

\noindent
\begin{minipage}{0.25\columnwidth}
    \vspace{1 mm}
    \includegraphics[width=\columnwidth]{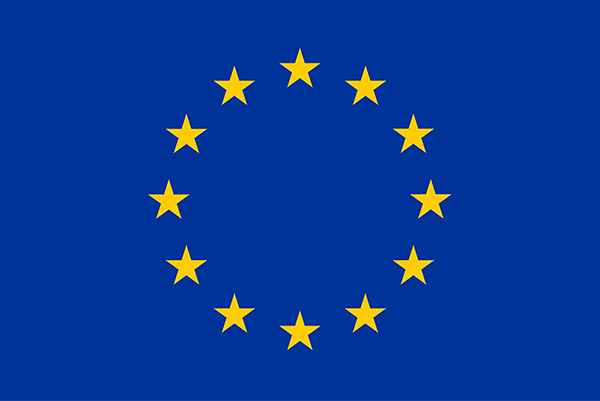}
\end{minipage} \hfill
\begin{minipage}{0.71\columnwidth}
    \vspace{1.5 mm}
    \footnotesize{This work has received funding from EU programmes Horizon 2020  (No.~861222 -- MINTS) and Horizon Europe (No. 101096302 -- 6GTandem and No.~101059091 -- TALENT), as well as from the Swedish Research Council (No.~2022-04691), ELLIIT, and Ericsson.}
\end{minipage}

\bibliographystyle{IEEEtran}
\bibliography{refs}

\newpage
\clearpage

\begin{wrapfigure}{l}{0.28\linewidth}
    \centering
    \includegraphics[width=1.08\linewidth]{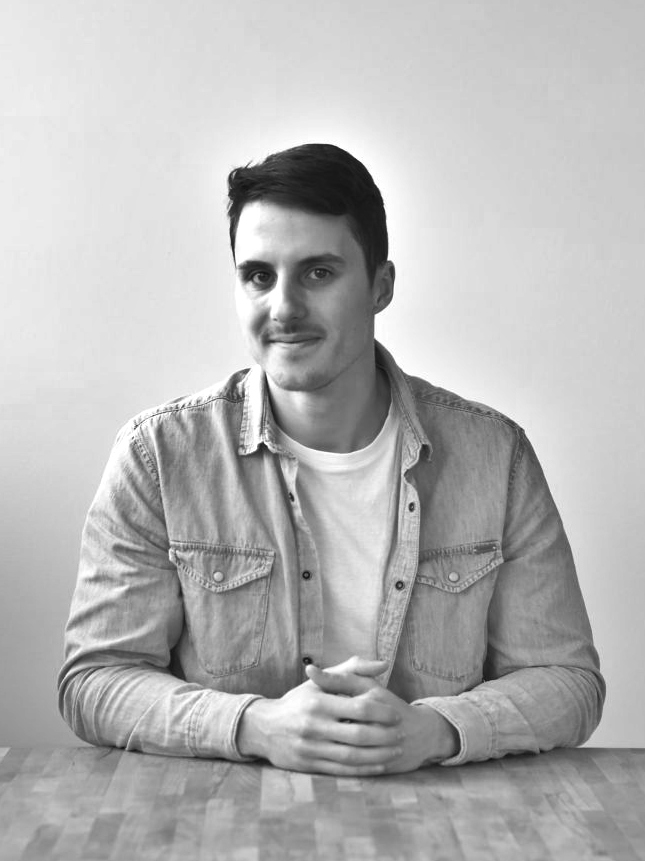}
    \vspace{-6mm}
\end{wrapfigure}
\noindent
\textbf{Alex Marinšek} (Student Member, IEEE) is a doctoral student at the DRAMCO lab at KU Leuven and part of the MINTS MSCA-ITN project, where he specializes in mmWave wireless communication for XR use cases. He received his M.Sc. degree in Electrical Engineering from the University of Ljubljana, Slovenia, where he was also a recipient of the national Zois scholarship. During his studies, he has authored and co-authored several scientific publications and designed an award-winning learning activity at the European Union's Science is Wonderful! 2023 competition. 

\vskip 2em

\begin{wrapfigure}{l}{0.28\linewidth}
    \centering
    \vspace{-3.6mm}
    \includegraphics[width=1.08\linewidth]{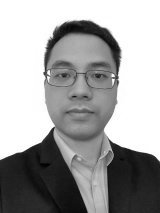}
    \vspace{-6mm}
\end{wrapfigure}
\noindent
\textbf{Xuesong Cai} (Senior Member, IEEE) received the B.S. degree and the Ph.D. degree (with distinction) from Tongji University, Shanghai, China, in 2013 and 2018, respectively. In 2015, he conducted a three-month internship with Huawei Technologies, Shanghai, China. He was also a Visiting Scholar with Universidad Politécnica de Madrid, Madrid, Spain in 2016. From 2018-2022, he conducted several postdoctoral stays at Aalborg University and Nokia Bell Labs, Denmark, and Lund University, Sweden. He is currently an Assistant Professor in Communications Engineering and a Marie Skłodowska-Curie Fellow at Lund University, closely cooperating with Ericsson and Sony. His research interests include radio propagation, high-resolution parameter estimation, over-the-air testing, resource optimization, and radio-based localization for 5G/B5G wireless systems. His work has led to over 64 peer-reviewed publications, 2 book chapters, and 4 granted patents.
    
Dr. Cai was a recipient of the China National Scholarship (the highest honor for Ph.D. Candidates) in 2016, the Outstanding Doctorate Graduate awarded by the Shanghai Municipal Education Commission in 2018, the Marie Skłodowska-Curie Actions (MSCA) “Seal of Excellence” in 2019, the EU MSCA Fellowship (ranking top 1.2\%, overall success rate 14\%) and the Starting Grant (success rate 12\%) funded by the Swedish Research Council in 2022. He was also selected by the “ZTE Blue Sword-Future Leaders Plan” in 2018 and the “Huawei Genius Youth Program” in 2021. He is an AP-S 2024 Young Professional Ambassador and serves as an Editor of IEEE Transactions on Vehicular Technology, IET Communications, and Wireless Communications and Mobile Computing.

\vskip 2em

\begin{wrapfigure}{l}{0.28\linewidth}
    \centering
    \vspace{-3.6mm}
    \includegraphics[width=1.08\linewidth]{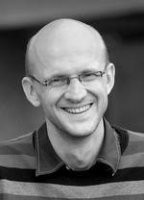}
    \vspace{-6mm}
\end{wrapfigure}
\noindent
\textbf{Lieven De Strycker} (Member, IEEE) is a full professor at the Faculty of Engineering Technology, Department of Electrical Engineering, KU Leuven. He was a coordinator of European Erasmus Intensive teaching programs (Life Long Learning program). Prof. L De Strycker was an invited speaker on indoor localization at several universities (Darmstadt, Porto, Iasi, Plovdiv, Suceava, La Rochelle). In 2001 he joined the Engineering Technology department of the Catholic University College Ghent (KAHO Sint-Lieven), where he founded the DRAMCO (wireless and mobile communications) research group in cooperation with ESAT-TELEMIC, KU Leuven. He is still the coordinator of the DRAMCO research group which has been involved in over 20 national and international research projects. He did an IAESTE traineeship at Siemens Madrid, Spain. He did research in European FP5 and national R\&D projects at the INTEC\_design Lab of Prof. Jan Vandewege and received the Ph.D. degree in Electrotechnical Engineering from Gent University in 2001 (summa cum laude). His Master degree in Electrotechnical Engineering he received at Ghent University, in 1996 (summa cum laude).

\vskip 20em

\begin{wrapfigure}{l}{0.28\linewidth}
    \centering
    \includegraphics[width=1.08\linewidth]{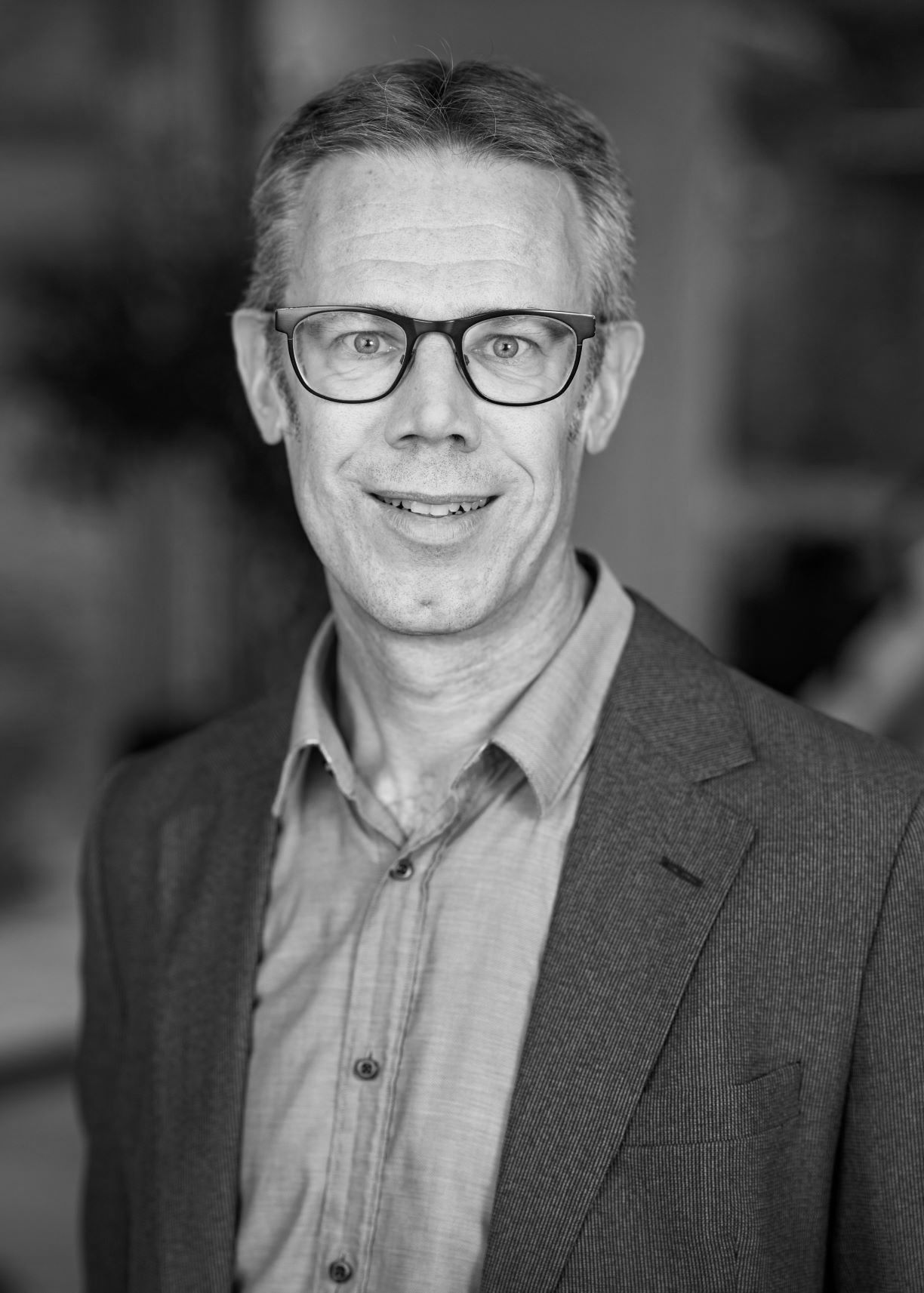}
    \vspace{-6mm}
\end{wrapfigure}
\noindent
\textbf{Fredrik Tufvesson} (Fellow, IEEE) received his Ph.D. in 2000 from Lund University in Sweden. After two years at a startup company, he joined the department of Electrical and Information Technology at Lund University, where he is now professor of radio systems. His main research interest is the interplay between the radio channel and the rest of the communication system with various applications in 5G/B5G systems such as massive MIMO, mm wave communication, vehicular communication and radio-based positioning. Fredrik has authored around 110 journal papers and 175 conference papers, he is fellow of the IEEE and his research has been awarded with the Neal Shepherd Memorial Award (2015) for the best propagation paper in IEEE Transactions on Vehicular Technology, the IEEE Communications Society best tutorial paper award (2018, 2021) and the IEEE Signal Proc. Society Donald G. Fink overview paper award 2023.

\vskip 4em

\begin{wrapfigure}{l}{0.28\linewidth}
    \centering
    \vspace{-3.6mm}
    \includegraphics[width=1.08\linewidth]{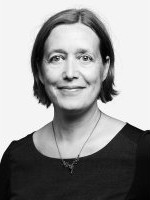}
    \vspace{-6mm}
\end{wrapfigure}
\noindent
\textbf{Liesbet Van der Perre} (Senior Member, IEEE) received the M.Sc. and Ph.D degree in Electrical Engineering from the KU Leuven, Belgium, in 1992 and 1997 respectively. Dr. Van der Perre joined imec’s wireless group in 1997 and took up responsibilities as senior researcher, system architect, project leader and program director, until 2015. She was appointed full Professor in the DRAMCO lab of the Electrical Engineering Department of the KU Leuven and guest professor at the University of Lund in 1996. Her main research interest is in energy efficient wireless connectivity and embedded systems, with applications in multiple antenna and large array systems, IoT and broadband networks.

\end{document}